\documentclass[fleqn,usenatbib]{mnras}

\usepackage{newtxtext,newtxmath}
\usepackage[T1]{fontenc}

\DeclareRobustCommand{\VAN}[3]{#2}
\let\VANthebibliography\thebibliography
\def\thebibliography{\DeclareRobustCommand{\VAN}[3]{##3}\VANthebibliography}

\usepackage{graphicx}	% Including figure files
\usepackage{amsmath}	% Advanced maths commands
\usepackage{gensymb}
\usepackage{xcolor}
\usepackage[normalem]{ulem}
\useunder{\uline}{\ul}{}
\usepackage{makecell}
\usepackage{rotating}
\usepackage{caption}
\usepackage{url}
\usepackage{hyperref}

\title[Gas detection around hybrid disc HD\,44892]{HD\,44892: The youngest (or oldest?) gas-harbouring debris disc around an intermediate mass star}

\author[K. M. Szewczyk et al.]{
Karolina M. Szewczyk$^{1}$,\thanks{E-mail: py19k2ms@leeds.ac.uk}
O. Panić$^{1}$,
D. P. Iglesias$^{1}$,
T. D. Pearce$^{2}$,
and J. M. Miley$^{3,4,5}$
\\
$^{1}$School of Physics \& Astronomy, University of Leeds, Sir William Henry Bragg Building, Woodhouse Lane, Leeds LS2 9JT, UK\\
$^{2}$Department of Physics, University of Warwick, Gibbet Hill Road, Coventry CV4 7AL, UK\\
$^{3}$Departamento de Física, Universidad de Santiago de Chile, Av. Victor Jara 3659, Santiago, Chile\\
$^{4}$Millennium Nucleus on Young Exoplanets and their Moons (YEMS)\\
$^{5}$Center for Interdisciplinary Research in Astrophysics Space Exploration (CIRAS), Universidad de Santiago de Chile, Chile
}

\date{Accepted XXX. Received YYY; in original form ZZZ}

\pubyear{\the\year{}}

\begin{document}
\label{firstpage}
\pagerange{\pageref{firstpage}--\pageref{lastpage}}
\maketitle

\begin{abstract}
We present the first detections of gas around a 2.1\,Myr old debris disc-bearing intermediate-mass star, HD\,44892. Gas is detected both in $^{12}$CO (2-1) emission through ALMA Band 6 observations and in absorption in Ca\,\textsc{ii} K and H, seen with high-resolution UVES spectroscopy. $^{13}$CO and C$^{18}$O (2-1) are not detected. The star exhibits a 12\,$\mu$m fractional excess of $7.86^{+0.11}_{-2.27}$, placing it in the transition stage between protoplanetary and debris discs. 
Our detection of 1.3\,mm emission yields the dust mass of 0.019\,$\pm$\,0.009\,M$_{\oplus}$ assuming 115\,K temperature. The $^{13}$CO non-detection places an upper limit on CO gas mass of $\sim$10$^{-2}$\,$M_{\oplus}$. The $^{12}$CO detection yields a CO gas mass of (7.86\,$\pm$\,2.05)\,$\times$\,10$^{-5}$\,$M_{\oplus}$ or (1.62\,$\pm$\,0.17)\,$\times$\,10$^{-4}$\,$M_{\oplus}$ assuming a gas temperature of 20\,K or 50\,K, respectively. These should be considered as lower limits since $^{12}$CO emission may be optically thick. With UVES, we find variability in Ca\,\textsc{ii} K and H lines, which can be interpreted by transiting circumstellar gas, ruling out interstellar absorption as their origin. Both the dust mass, which is within an order of magnitude of HD\,141569, and the gas mass derived here indicate a late gas dispersal stage of the protoplanetary disc. Through our analysis we deem the alternative age of 800\,Myr unlikely.
\end{abstract}


%%%%%%%%%%%%%%%%%%%%%%%%%%%%%%%%%%%%%%%%%%%%%%%%%%

%%%%%%%%%%%%%%%%% BODY OF PAPER %%%%%%%%%%%%%%%%%%

\section{Introduction}

While the dust content of debris discs has been extensively studied, less is known about the presence of gas. Numerous papers have searched for and examined this gas presence \citep[e.g.,][]{Rebollido2022The36546,Moor2017MolecularStars,Cataldi2023PrimordialALMA,Hales2022ALMADisk,Marino2016ExocometaryRing,Moor2019New32297, Iglesias2018DebrisOrigin}; however, many questions remain unanswered. Perhaps the biggest question of all is: where does this gas come from? Two main scenarios have been proposed. This gas is either primordial, leftover from the parent protoplanetary disc, or secondary, produced alongside dust, through processes such as planetesimal collisions, volatile sublimation, and photodesorption \citep{Hughes2018DebrisVariability}. Recent findings regarding the shielding effects in discs suggest that CO gas may accumulate in relatively high amounts from secondary means alone \citep{Kral2019ImagingDiscs}. Nonetheless, the primordial origin has not yet been disproved, especially as the effects of gas mixing complicate the case for the secondary gas scenario, making shielding less efficient \citep{Marino2022VerticalLifetime}, and models of \citet{Nakatani2021PhotoevaporationRemnants} suggest that, in case of A-type stars, the gas can survive as a protoplanetary remnant. 

\begin{figure}
    \centering
    \includegraphics[width=0.95\columnwidth]{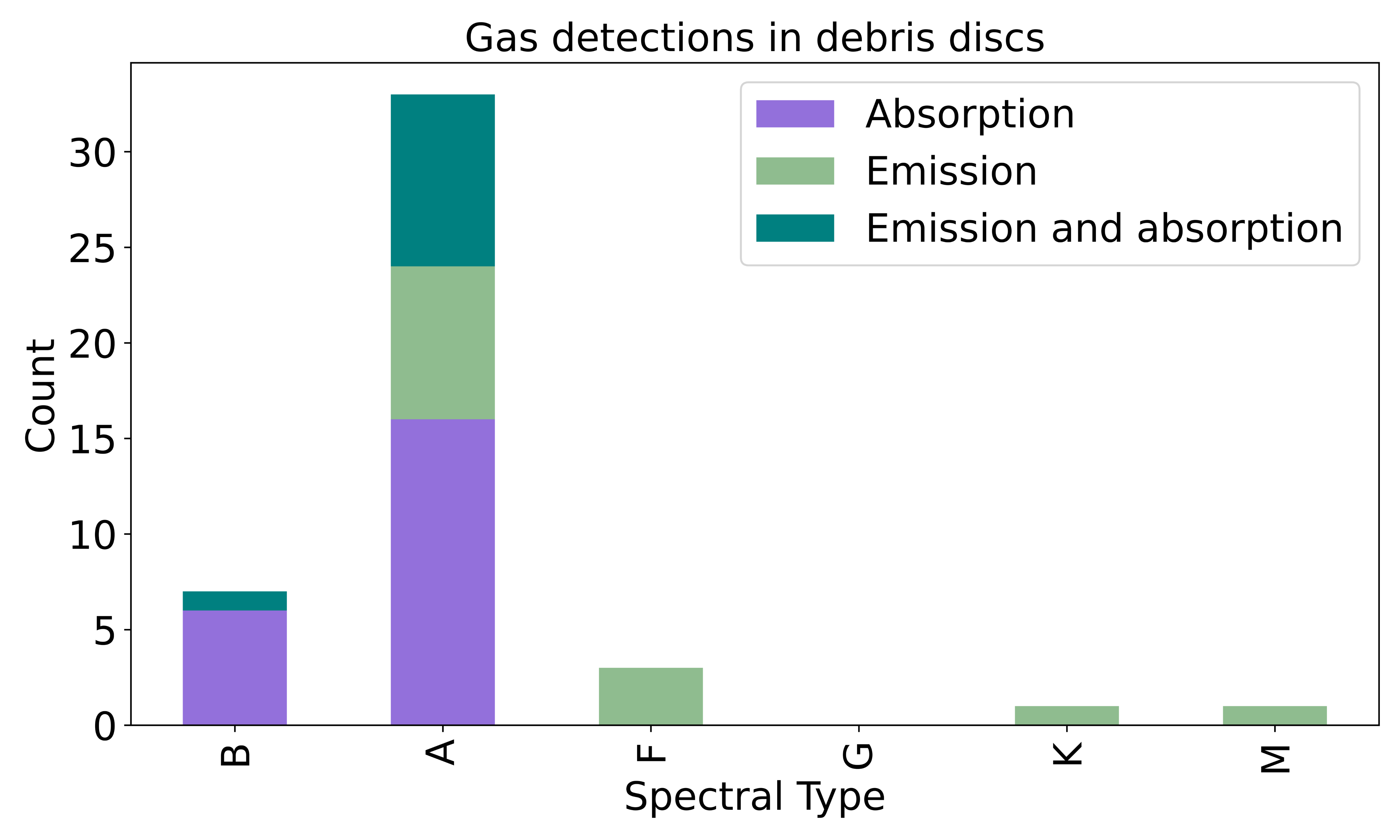}
    \caption{Histogram of all gas detections in debris discs through either emission or absorption, divided into bins of stellar type. For a full list of objects with gas emission, see Table\,\ref{tab:allCOGasDetections}. Objects with gas absorption used in this figure were adopted from \citet{Iglesias2020SearchingDisks} (Table\,1.1), based on data from \citet{Chen2003TheHerculis}, \citet{Lagrange-Henri1990SearchStar.}, \citet{Abt1973RotationDwarfs.}, \citet{Welsh2013CircumstellarExocomets}, \citet{Koubsky1993ComingHerculis.}, \citet{Iglesias2018DebrisOrigin}, \citet{Iglesias2019AnDisc}, \citet{Welsh1998Beta85905}, \citet{Welsh2015TheAbsorption}, \citet{Montgomery2012DetectionStars}, \citet{Welsh2018FurtherDiscs}.}
    \label{fig:histogramAllDetections}
\end{figure}

As of August 2025, 45 debris discs containing gas have been discovered either through gas emission or absorption. Over 20 of them have been observed with ALMA to search for CO and other gas emission, a full list of which can be found in Table\,\ref{tab:allCOGasDetections}. Interestingly, the majority of all gas detections in debris discs have been around A-type stars (see Figure\,\ref{fig:histogramAllDetections}). For most of these systems, secondary origin models can readily explain the observed low CO gas masses. However, a few systems, such as 49\,Cet \citep[45\,Myr,][]{Moor2019New32297, Pearce2022PlanetDiscs}, HD\,21997 \citep[42\,Myr,][]{Kospal2013ALMA21997,Moor2013ALMA21997}, HD\,141569 \citep[5\,Myr,][]{Miley2018Unlocking141569}, and HD\,32297 \citep[<\,30\,Myr,][]{Moor2019New32297}, which are intermediate-mass stars, exhibit significantly higher CO gas masses, which the current secondary models are not able to fully explain. In these cases, the origin of the gas is proposed to be primordial or a mixture of primordial and secondary \citep{Cataldi2023PrimordialALMA}. Secondary origin is most certain around older systems, such as the 440\,Myr Fomalhaut \citep{Matra2017ExocometaryObservations} or the 1-2\,Gyr $\eta$\,Crv \citep{Marino2017ALMAPlanets}, where primordial gas cannot survive over such long timescales. $\beta$\,Pic, despite its younger age of $\sim$20\,Myr, is another example of a system with secondary gas, in which gas produced in planetesimal collisions and the destruction of volatile bodies has been observed \citep{Dent2014MolecularDisk,Matra2017DetectionComets}.

\begin{figure*}
    \centering
    \includegraphics[width=2\columnwidth]{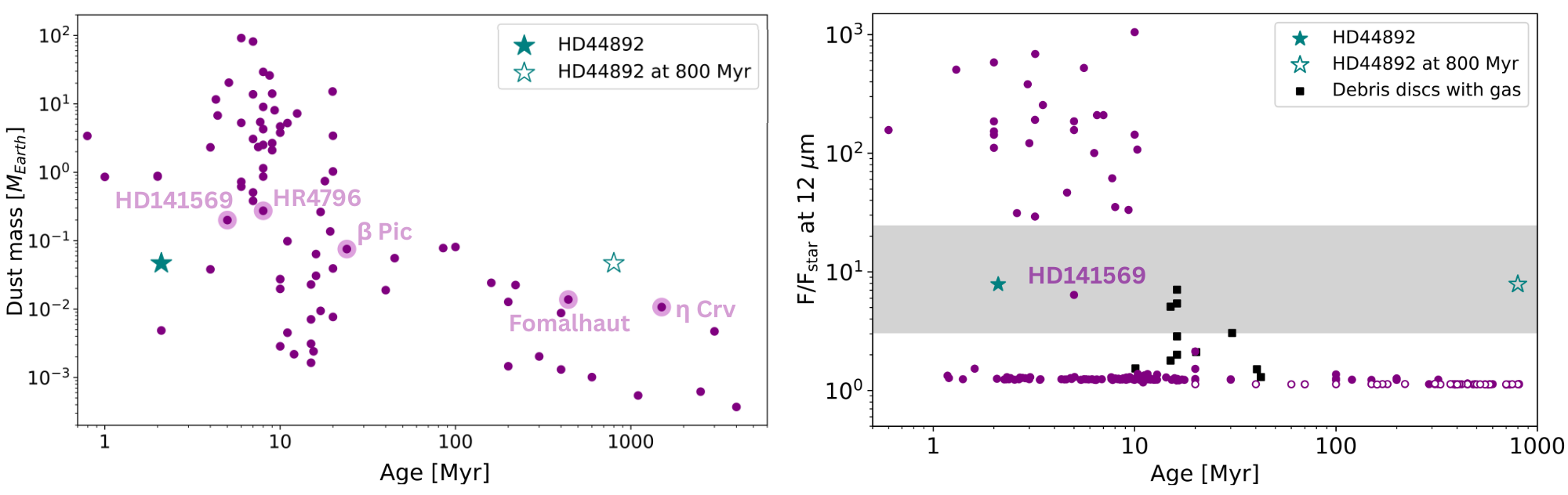}
    \caption{\textbf{Left: }Evolution of dust mass in debris discs, originally presented in \citet{Wyatt2015FiveDisk}. We have added more dust masses of debris discs, including those with gas discovered since. HD\,44892 is marked with the two different possible ages: 2.1\,Myr and 800\,Myr. At the older age HD\,44892 would have an unexpectedly high dust mass compared to other, older debris discs. Continuum fluxes used to create this plot were adopted from \citet{Panic2013FirstDiscs,Sandell2011ASTARS,Moor2017MolecularStars,Sheret2004SubmillimetreStars,Lieman-Sifry2016DebrisALMA,Pearce2022PlanetDiscs,Stapper2022TheALMA}, excluding systems with highly uncertain age estimates. The age of the HR\,4796 was taken from \citet{Rhee2007CharacterizationCatalogs}. Dust temperature of 50\,K is assumed for debris discs, and 25\,K for protoplanetary discs, with a dust opacity assumed as $\kappa=0.1\times2.3\times(\frac{\nu}{230\times10^9})^{0.7}$ [cm$^{2}$\,g$^{-1}$] \citep{Beckwith1990}, where $\nu$ is the frequency of observations [Hz]. \textbf{Right: }IR excess measured at 12\,$\mu$m plotted as a function of age of the system, as originally presented in \citet{Wyatt2015FiveDisk} (in purple, where open circles are upper limits), with additional data from \citet{Moor2017MolecularStars} of gas-bearing debris discs (in black). HD\,44892 is again marked with the two possible ages, with the older age making it an outlier in the trend. The grey-shaded region marks the stage between protoplanetary and debris discs, where hybrid discs are expected to reside.}
    \label{fig:DustAgeEvolution}
\end{figure*}

The primordial origin has been suggested for discs where the gas is present in substantial amounts - quantities that are not yet fully explainable by models of secondary origin - and where the parent star is very young \citep[primordial gas is possible in systems <\,50\,Myr-old in models of][]{Nakatani2021PhotoevaporationRemnants}, hence the gas might not have yet been fully cleared from the system \citep{Smirnov-Pinchukov2022LackGas}. Such is the case for HD\,141569 \citep{Miley2018Unlocking141569}, a 5\,Myr star harbouring a `hybrid' disc, name used for gas-bearing discs with protoplanetary disc levels of likely primordial gas, but debris-disc levels of secondary dust. These discs are similar to dusty debris discs but retain some primordial gas from the earlier evolutionary stage, hence are hybrid between protoplanetary and debris discs \citep{Pericaud2017TheDisks}. The dissipation of primordial gas is still ongoing in these systems, placing them just at the end of protoplanetary disc dispersal. In the case of HD\,141569, the dust is of secondary origin or mixed, with a relatively low dust mass and IR excess for a protoplanetary disc but high for a debris disc (Figure\,\ref{fig:DustAgeEvolution}), while the gas is suspected to be primordial and at a mass comparable to the least-massive protoplanetary discs (PPDs). HD\,141569 is thus the only known gas-harbouring debris disc found around a pre-main-sequence (pre-MS) intermediate-mass star (IMS) with a measured CO mass to date. 

\section[HD 44892]{HD\,44892}

In this paper, we present ALMA observations of HD\,44892 (also known as HIP\,30414), an A9/F0IV type star \citep[SIMBAD,][]{Wenger2000TheDatabase} located 191\,$\pm$\,1\,pc away \citep{Vallenari2023iGaia/i3,Prusti2016TheMission}. HD\,44892 has a temperature of 7500\,$\pm$\,500K, a luminosity of 60.0\,$\pm$\,20.0\,$L_{\odot}$, stellar mass of 2.87\,$^{+0.67}_{-0.41}$\,$M_{\odot}$, and an age of 2.1\,$^{+1.2}_{-1.0}$\,Myr \citep{Iglesias2023X-shooterEvolution}. The pre-MS star HD\,44892 has undergone the transition from convective to radiative structure at $\sim$1\,Myr. \citet{Gontcharov2006PulkovoSystem} have calculated the radial velocity of the star to be 20.9\,$\pm$\,0.6\,km\,s$^{-1}$, which is the value adopted in this study. A more recent estimate yields a value of 20.88\,$\pm$\,1.85\,km\,s$^{-1}$ (D. Iglesias, priv. comm.), which is consistent with previous findings. There are no mentions in the literature suggesting HD\,44892 is part of a binary or multiple star system; it is also a field star \citep{Gagne2018BANYAN.Pc}. 

\citet{Ishihara2017FaintIRSF} identified it as a warm debris disc candidate based on its IR excess at 18\,$\mu$m found through spectral energy distribution (SED) fitting, and it was first identified as a hybrid disc candidate in \citet{Iglesias2023X-shooterEvolution}. HD\,44892 has a fractional IR excess at 12\,$\mu$m of 7.86\,$^{+0.11}_{-2.27}$ \citep{Iglesias2023X-shooterEvolution}, placing it between debris ($3<F_{12\mu m}/F_{star}$) and protoplanetary ($F_{12\mu m}/F_{star}>25$) discs \citep{Wyatt2015FiveDisk, Iglesias2023X-shooterEvolution}, where also a number of gas-bearing debris discs have been found (Figure\,\ref{fig:DustAgeEvolution}, right).

There is a disagreement in the literature regarding the stellar parameters of HD\,44892, particularly about its age. Its location on the HR diagram can be fit by either pre- or post-MS tracks. Past studies, such as \citet{Zorec2012RotationalStars} and \citet{Bochanski2018FundamentalGaia}, have classified it as a post-MS object ($\sim$570-580\,Myr with a mass of 2.50-2.60\,$M_{\odot}$). \citet{Iglesias2023X-shooterEvolution} present the fit to the pre-MS tracks yielding the 2.1\,Myr and an alternative fit to the post-MS tracks with 800\,Myr, with a mass of 2.29\,$M_{\odot}$. They conclude that the possibility of the older age is less likely. In this paper, we adopt the pre-MS age of 2.1\,Myr and present arguments against the alternative age of 800\,Myr in later sections.

\section{ALMA}

\subsection{ALMA observations and data reduction}

HD\,44892 was observed using the ALMA (Atacama Large Millimeter/submillimeter Array) 12\,m array on 15th January 2023 (PI Panić, ID 2022.1.01686.S). Observations were taken across four spectral windows (SPWs) in Band 6, searching for both continuum and gas emission. The target was observed for a total of 34 minutes and 25 seconds using 46 operational antennas, with minimum and maximum baselines of 15.3\,m and 783.5\,m, respectively. J0423-0120 was used as the flux calibrator and J0609-1542 as the phase calibrator.

Band 6 transitions of CO were observed, including the 2-1 transitions of ${}^{12}\text{CO}$ in a SPW of 1920 channels of width 244.141\,kHz and a central frequency of 230\,517\,MHz, resulting in a resolution of 0.32\,km\,s$^{-1}$. ${}^{13}\text{CO}$ and $\text{C}^{18}\text{O}$ SPW has 1920 channels of width 976.562\,kHz and a central frequency of 219\,980\,MHz, which gives a spectral resolution of 1.33\,km\,s$^{-1}$. The remaining two spectral windows, of central frequencies 216\,980\,MHz and 232\,979\,MHz, were dedicated to continuum emission. Each of these windows contains 128 channels of width 15\,625\,kHz.

We processed all of the data using \textsc{CASA} 6.5.4-9 \citep{McMullin2007CASAApplications} and \textsc{CARTA} 4.0.0 \citep{Comrie2018CARTA:Astronomy}. Continuum emission of HD\,44892 was imaged using the task `tclean' with Briggs weighting and a robust parameter set to 0.3, which we found to provide the highest signal-to-noise ratio and source emission over the range of -2 to 2 explored. Our target's emission fits entirely in the region where primary beam is maximum, hence no primary beam correction was applied. No self-calibration was applied to the data as the source is too weak in continuum to provide reliable phase or amplitude solutions. 

For line data, we used the \textsc{CASA} task `uvcontsub' to subtract the continuum from the line emission data. Using the `plotms' task (amplitude vs channel), we identified SPWs and frequencies at which line emission was potentially present. We imaged both SPWs using the `tclean' task with Briggs weighting and a robust parameter set to 0.3, which produced the highest signal-to-noise ratio for the $^{12}$CO detection. We did not detect any emission in the individual channels of the SPW containing ${}^{13}$CO and C$^{18}$O. 

\subsection{ALMA results}

\subsubsection{Continuum}\label{sec:dustFlux}

\begin{figure}
    \centering
    \includegraphics[width=\columnwidth]{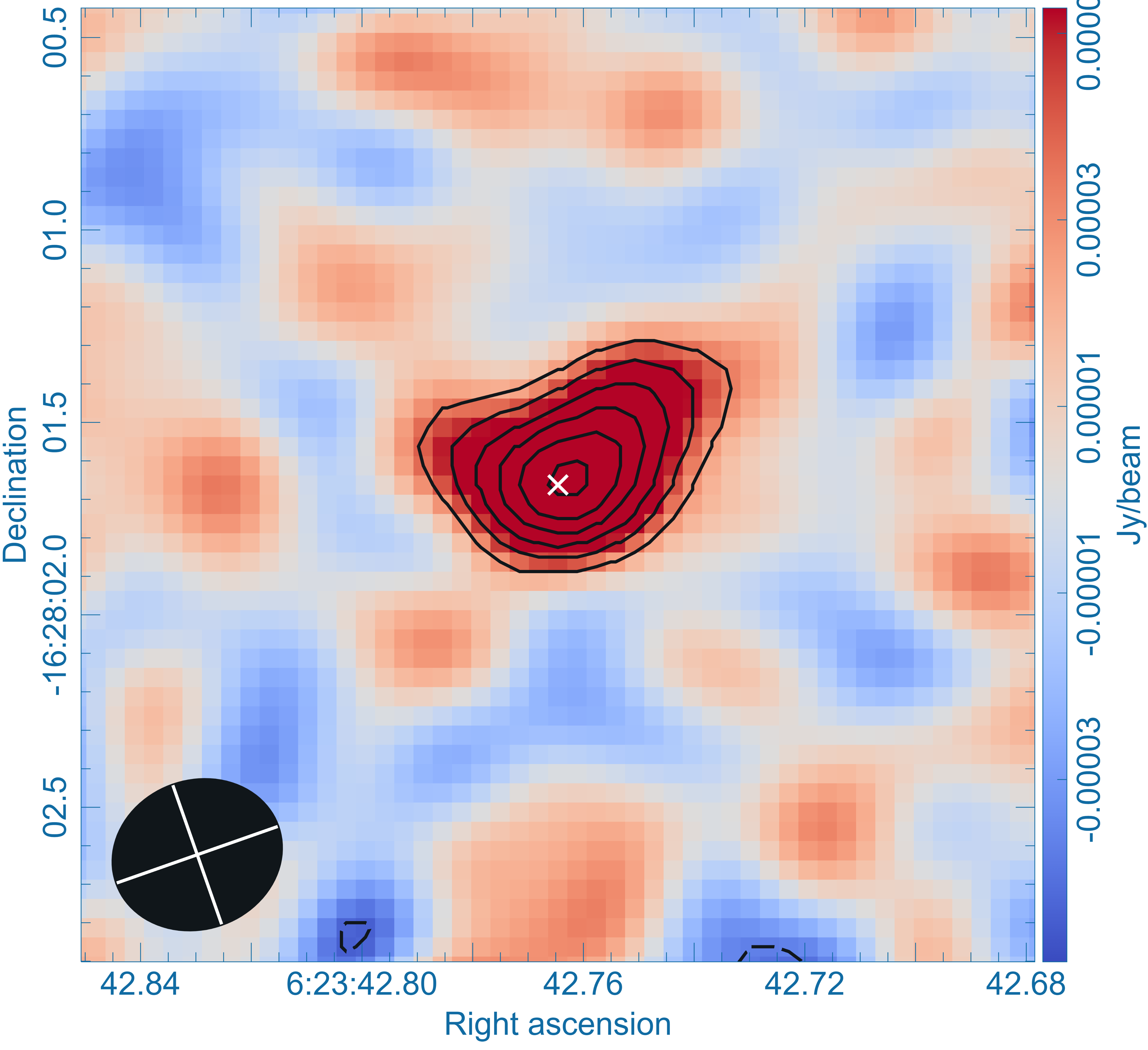}
    \caption{Continuum emission of HD\,44892, marked in contours of (\,$\pm$\,2, 3, 4, 5, 6, 7)\,$\times$\,$\sigma$, where $\sigma=0.016$\,mJy\,beam$^{-1}$. Position of HD\,44892 marked as a white cross.}
    \label{fig:HD44892dust}
\end{figure}

\begin{figure}
    \centering
    \includegraphics[width=0.95\columnwidth]{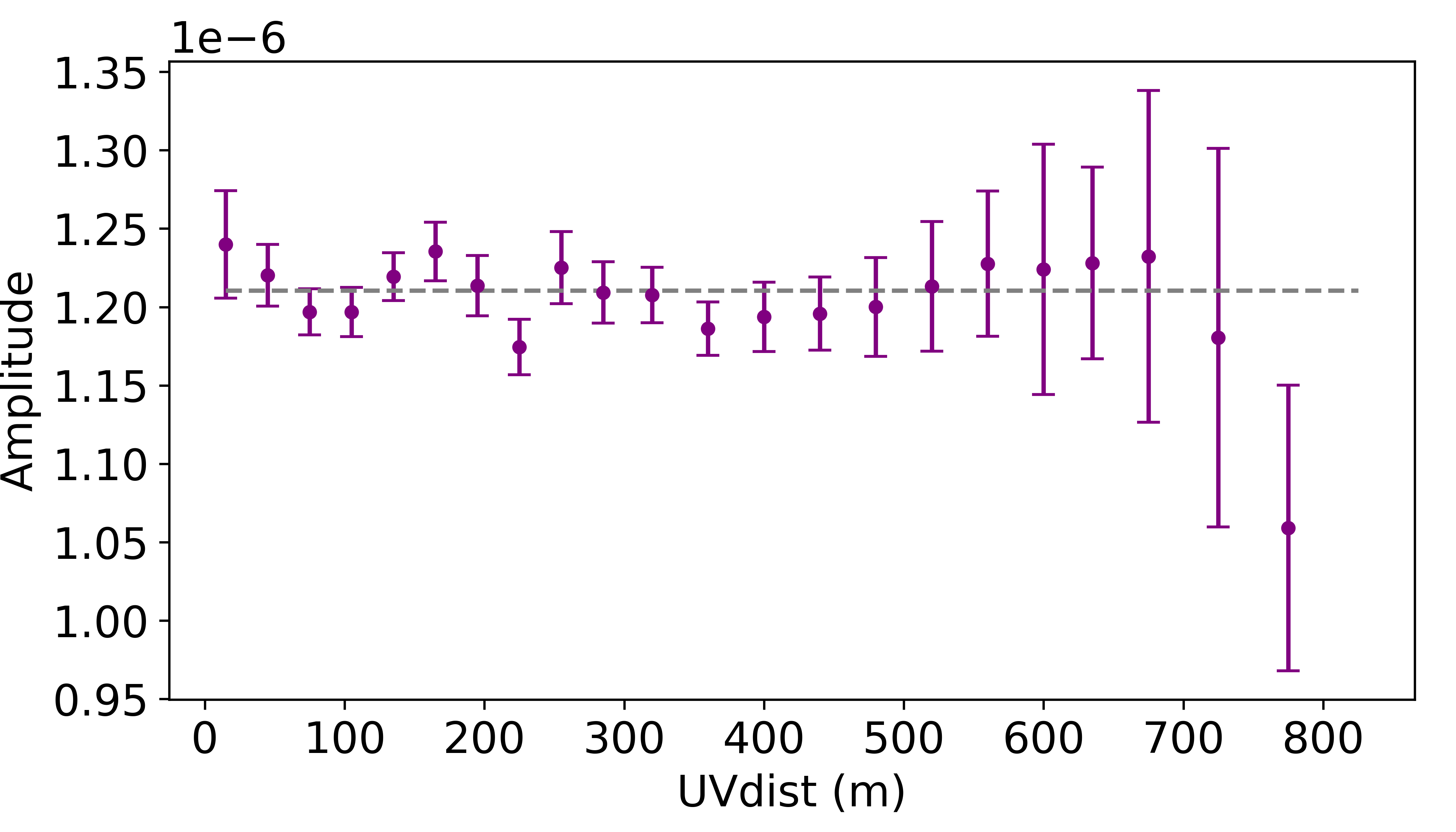}
    \caption{Amplitude vs UVdist plot of continuum of HD\,44892. While the shorter UV-distance baselines can be fit with a straight line (dashed-grey), hence are unresolved, the longest UV-baselines cannot, hence are not unresolved and deviate from a point-like source. The few points shown here correspond to the full continuum dataset that has been averaged in time, channels, and binned by UVdist for clarity.}
    \label{fig:ContMargResolved}
\end{figure}

For continuum of HD\,44892 our data reduction resulted in a synthesised beam size of 0\farcs44\,(84\,au)\,$\times$\,0\farcs38\,(72\,au) at a position angle (PA) of $-70.7\degree$. We detected continuum emission of 
0.136\,$\pm$\,0.016\,mJy ($\sim$9\,$\sigma$), shown in Figure\,\ref{fig:HD44892dust}. The continuum flux was measured only from inside the 3\,$\sigma$ contour. Using task `imfit', which is part of the \textsc{CASA} package, we found that the continuum emission of HD\,44892 is marginally resolved, as the fitted Gaussian component has a FWHM of major axis of 0\farcs58\,$\pm$\,0\farcs08, 110\,au, which extends beyond that of the beam (0\farcs43, 82\,au, measured in the same direction). The drop in amplitude at longest UV distances (Figure\,\ref{fig:ContMargResolved}), confirms that the most extended emission from the source is being resolved.

The continuum emission appears to be offset from the position of HD 44892 (see Figure\,\ref{fig:HD44892dust}). To determine the underlying cause of this apparent offset, we calculated the astrometric accuracy of ALMA\footnote{\url{https://help.almascience.org/kb/articles/what-is-the-absolute-astrometric-accuracy-of-alma}} using Equation\,\ref{eq:PosAcc}:

\begin{equation} \label{eq:PosAcc}
%pos_{acc} = \frac{beam_{FWHP}}{0.9 \times SNR}
\sigma_{\mathrm{pos}} = \frac{\theta_{\mathrm{FWHM}}}{0.9 \times \mathrm{SNR}}
\end{equation}

where $\theta_{\mathrm{FWHM}}$ is the full width at half maximum (FWHM) of the synthesized beam, and SNR is the signal-to-noise ratio of the peak target flux. For our data we found $\sigma_{\mathrm{pos}}$\,=\,45\,milliarcseconds. The apparent offset of the continuum emission from the stellar position is 34\,milliarcseconds, which is less than the calculated $\sigma_{\mathrm{pos}}$, suggesting that it is most likely due to the limited astrometric accuracy rather than physical displacement.

\subsubsection{Gas}

\begin{figure}
    \centering
    \includegraphics[width=1\columnwidth]{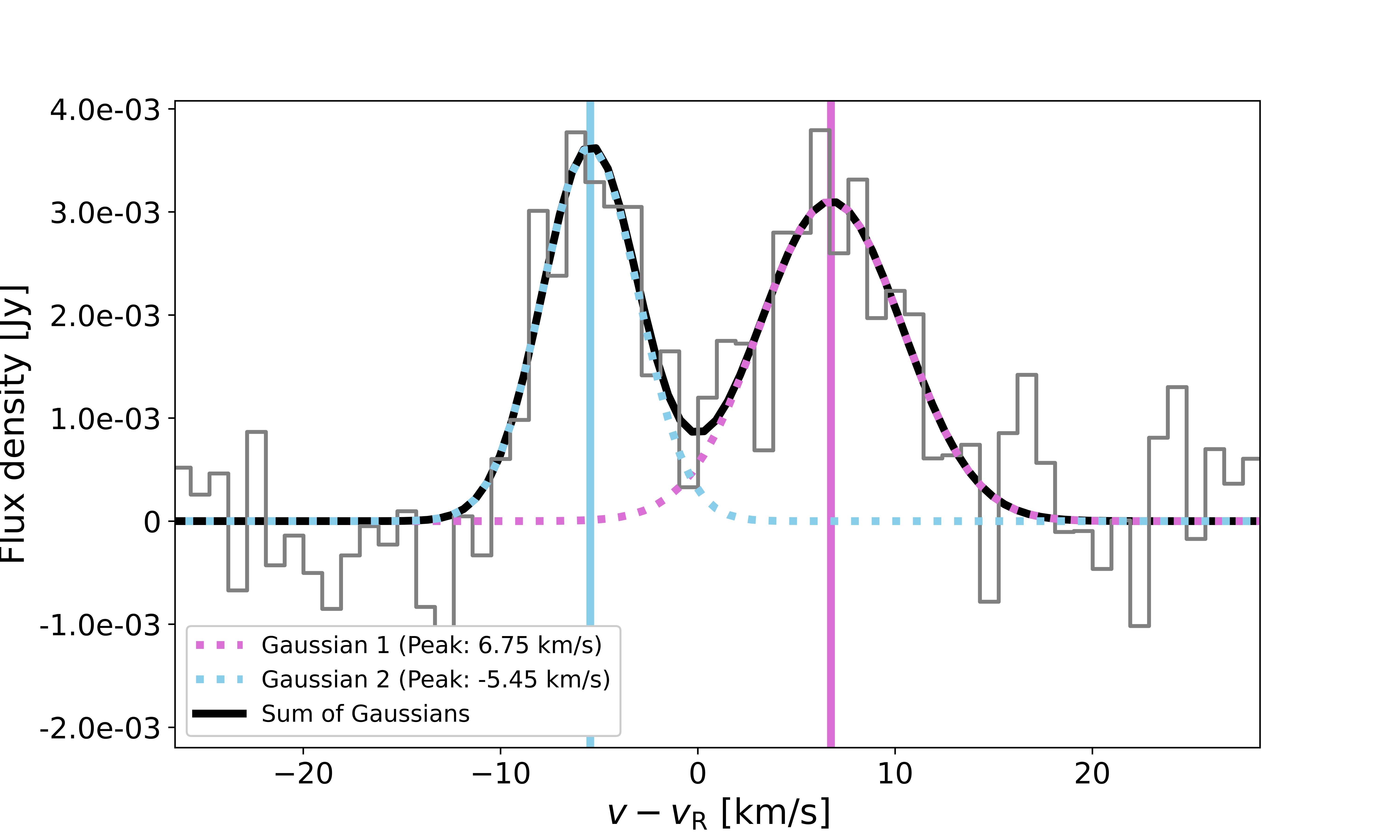}
    \caption{Radial velocity spectrum of $^{12}$CO emission with peak positions marked. The x-axis has been corrected for the star's radial velocity in the LSR frame of $\sim$1.5\,km\,s$^{-1}$.}
    \label{fig:HD44892velStructure}
\end{figure}

For ${}^{12}$CO emission, the resulting synthesized beam size is 0\farcs46 (88\,au)\,$\times$\,0\farcs40 (76\,au), at a PA of -80.6$\degree$. Figure\,\ref{fig:HD44892velStructure} shows the spectrum of the detected gas emission, which has the typical double-peaked profile associated with gas rotating in a Keplerian disc. Fitting a double-peaked Gaussian curve to this spectrum reveals a midpoint at $v={\sim}0$\,km\,s$^{-1}$. The first peak is at -5.45\,km\,s$^{-1}$ and the second at 6.75\,km\,s$^{-1}$. This x-axis of this plot has been corrected for the star's radial velocity in the LSR frame of ${\sim}1.5km\,s^{-1}$. 

\begin{figure}
    \centering
    \includegraphics[width=\columnwidth]{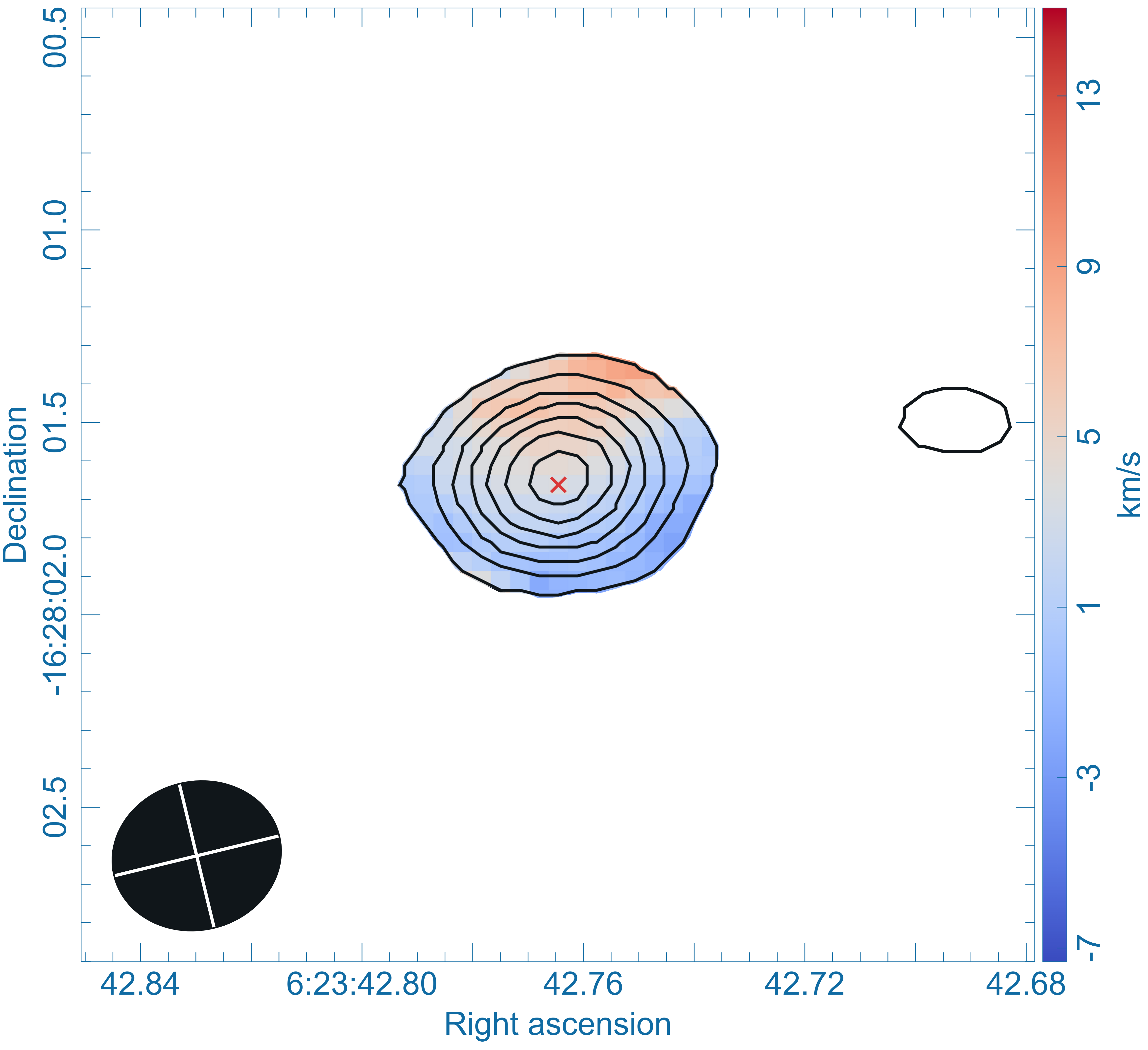}
    \caption{Integrated $^{12}$CO 2-1 line intensity of HD\,44892 with contours of (2, 3, 4, 5, 6, 7, 8)\,$\times$\,$\sigma$, where $\sigma=11.01$\,mJy\,beam$^{-1}$\,km\,s${}^{-1}$. Position of HD\,44892 marked as red cross. Contours are integrated intensity (moment 0) while colour scale is from intensity-weighted coordinate (moment 1), only shown for $>2\sigma$ emission from the source.}
    \label{fig:HD44892gas}
\end{figure}

From moment 0 (Figure\,\ref{fig:HD44892gas}, integrated intensity made using the \textsc{CASA} task `immoments' within range -8.49-15.33\,km\,s$^{-1}$) we found an integrated line flux of 96.92\,$\pm$16.22\,mJy\,km\,s$^{-1}$ ($\sim$6.0\,$\sigma$). Therefore, we decided to re-image the data, this time binning three channels together during tclean (nchan\,=\,3). This reduced the noise in our image and made the emission more prominent. The moment 0 map was made using the same velocity range as in the non-binned case, resulting in a slightly lower noise of 14.10\,mJy\,km\,s$^{-1}$, and the integrated line flux of 93.06\,mJy\,km\,s$^{-1}$, improving the signal-to-noise (6.6\,$\sigma$). 

\begin{figure*}
    \centering
    \includegraphics[width=2\columnwidth]{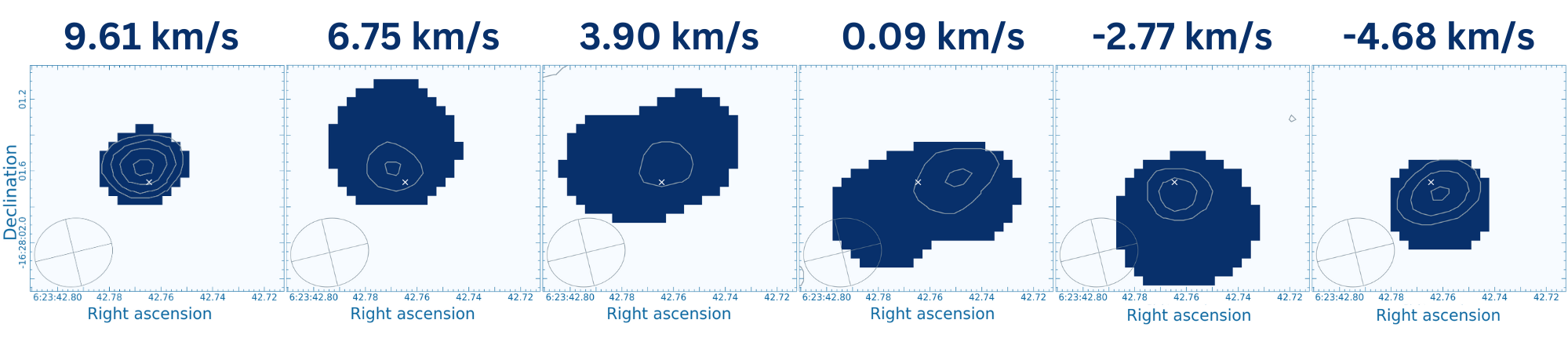}
    \caption{Contours of $^{12}$CO emission of HD\,44892 overlaid on top of the created Keplerian Mask. The gas emission rotates with the theoretical Keplerian motion around the host star. Channel contours start at 3\,$\sigma$ ($\sigma=1.60$ mJy beam$^{-1}$) with the exception of middle two panels which start at 2\,$\sigma$.}
    \label{fig:KeplerianMask}
\end{figure*}

We then also used the Keplerian Mask tool\footnote{\url{https://github.com/richteague/keplerian_mask}}, which produces a rotating mask based on the system's parameters that a Keplerian disc should follow. To search the parameter space to be used in the tool, we are guided by the aspect ratio of the deconvolved disc size. We used the `imfit' task and obtained the deconvolved Gaussian parameters: 0\farcs34\,$\pm$\,0\farcs10 for the major axis and 0\farcs19\,$\pm$\,0\farcs08 for the minor axis (PA of 76.2\,$\pm$\,63.4$\degree$). This suggests an inclination of 56.1\,$\pm$\,19.9$\degree$. We therefore varied the inclination from 36$\degree$ to 76$\degree$ in the Keplerian Mask tool.
The stellar mass range was adopted from \citet{Iglesias2023X-shooterEvolution} as 2.87\,$^{+0.67}_{-0.41}M_{\odot}$ and Keplerian rotation computed accordingly in the tool. Using these values as initial constraints, we explored combinations of the parameters (within their uncertainty ranges) to best match the $^{12}$CO emission of HD\,44892. The parameters that resulted in the best fit were an inclination of 40$\degree$, PA of 17.2$\degree$, stellar mass of 3.27\,$M_{\odot}$, and an outer radius of 0\farcs4 (76\,au). The $^{12}$CO emission follows this rotation, indicating that the emission lines up with theoretical positions and confirming it is from a disc rotating around the parent star (Figure\,\ref{fig:KeplerianMask}). Using the Keplerian Mask during cleaning has improved the signal-to-noise once more, in moment 0 this resulted in an integrated line flux ($F_{21}$) of 95.47\,$\pm$\,13.49\,mJy\,km\,s$^{-1}$ (7.1\,$\sigma$, Figure\,\ref{fig:HD44892gas}). Here we used channels where gas emission was most prominent, corresponding to the frequency range of 230.5262\,GHz (15.33\,km\,s$^{-1}$) to 230.5445\,GHz (-8.49\,km\,s$^{-1}$).  

Knowing the channels (velocities) at which $^{12}$CO emission is present, we created a moment 0 map for the SPW containing $^{13}$CO and C$^{18}$O. The emission from these isotopes would allow for a more precise CO gas mass calculation, as they are likely optically thin, ensuring the $^{12}$CO emission traces molecular density, rather than being convoluted with temperature effects. Since no emission was visible in the individual channels of this spectral window, we used the same velocity range as for $^{12}$CO. However, this has resulted in a non-detection of both $^{13}$CO and C$^{18}$O.

Figure\,\ref{fig:HD44892gas} shows the full extent of ${}^{12}$CO gas emission (contour levels) overlaid on the velocity distribution of the disc. The visible velocity gradient, from red- to blue-shifted, indicates gas rotation. This, along with the disc being located around the star's position, confirms the gas is rotating around the star and the emission is not from a different source. The gas emission appears compact and follows the beam shape, with the highest intensity near the centre, as expected for a spatially unresolved source.

\section{UVES}

\begin{figure*}
    \centering
    \includegraphics[width=2\columnwidth]{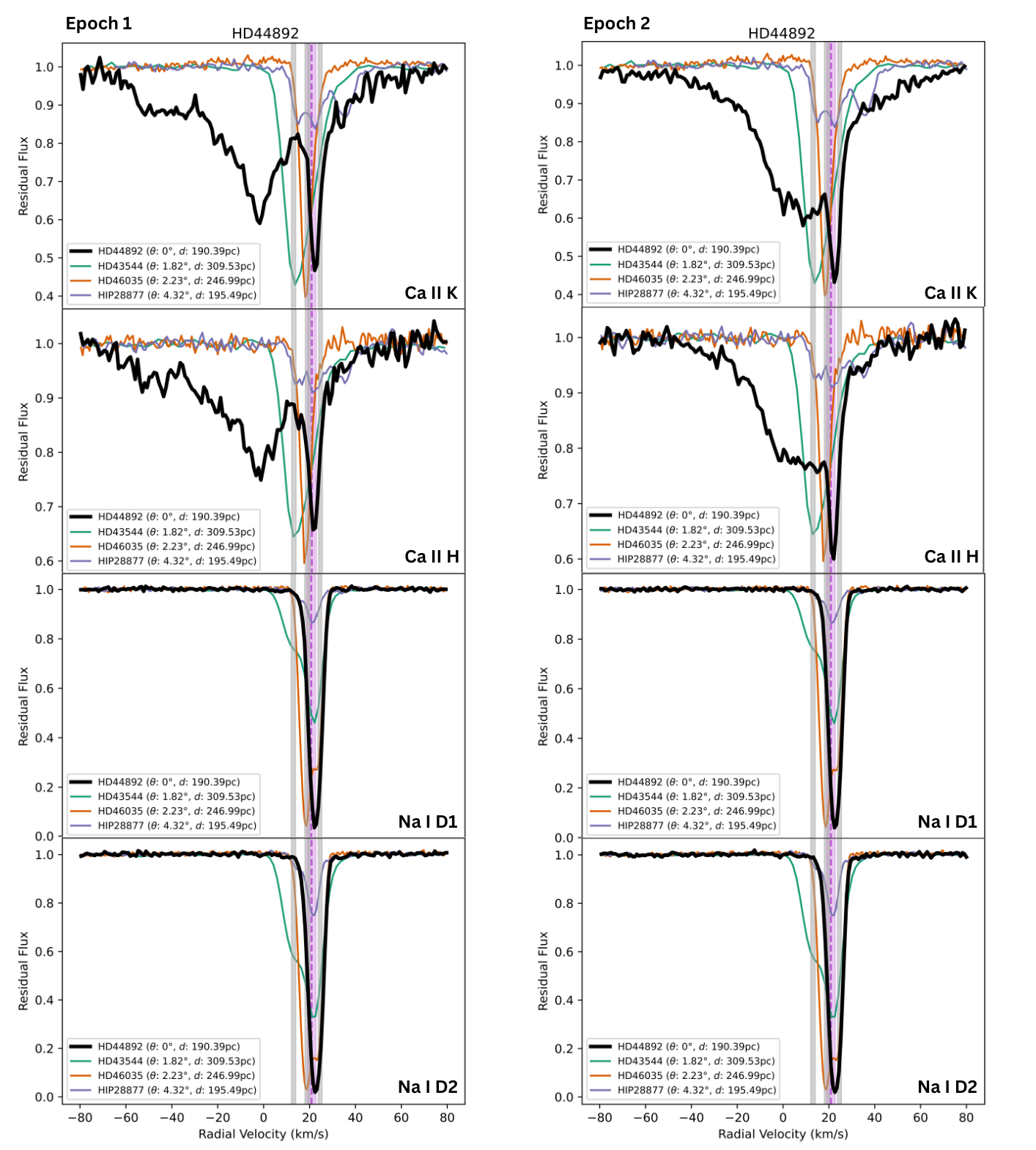}
    \caption{Narrow absorption features found in the spectra of HD\,44892 shown in black, overlaid with features found in the spectra of nearby stars, with their distances and separations from HD\,44892 marked. Each row represents a different absorption line: Ca\,\textsc{ii} K \& H, and Na\,\textsc{i} D1 \& D2. The two columns show observations from two different epochs. Grey-shaded regions correspond to the radial velocities of known clouds in the line of sight, while the pink region is the radial velocity of HD\,44892, 20.88\,$\pm$\,1.85\,km\,s$^{-1}$. Variability between the two epochs is observed in the calcium (K and H) lines but is absent in the sodium (D1 and D2) lines, spawning a radial velocity range of -70 to 15\,km\,s$^{-1}$. This variability may be caused by transiting, circumstellar gas. The pronounced narrow absorption feature at ~22.5\,km\,s$^{-1}$ is most likely caused by interstellar gas, given its strength in both sodium lines and the close match with nearby stars' features in calcium lines.}
    \label{fig:AbsAndSkyMap}
\end{figure*}

\subsection{UVES observations and data reduction}

We obtained high-resolution observations of HD\,44892 using UVES (Ultraviolet and Visual Echelle Spectrograph; \citealt{Dekker2000DesignObservatory}) at the VLT (Very Large Telescope), on 13th and 27th September 2023 (PI Iglesias, ID 112.262P.001). VLT is part of the European Southern Observatory and is located in Cerro Paranal in Chile. Data was taken in both Blue and Red arms, centred at 390\,nm and 580\,nm, respectively. We used the 0\farcs5 for these observations to achieve high resolution (R$\sim$74000 for the Red arm and R$\sim$65000 for the Blue arm). HD\,44892 was observed each time for 1 minute 48 seconds. Data of one of the comparison stars used, HIP\,28877, is part of the same observing programme. The observations were reduced using the UVES EsoReflex pipeline \citep{Ballester2000ThePipeline, Freudling2013AutomatedAstronomy} version 6.4.1.

Additional spectra from the ESO archive\footnote{\url{http://archive.eso.org}} were obtained for two other comparison stars (HD\,43544, HD\,46035), using high-resolution spectrographs HARPS \citep{Mayor2003SettingHARPS} and FEROS \citet{Kaufer1999CommissioningLa-Silla.} at the La Silla Observatory. These were science-ready products from the ESO archive and reduced with each instrument standard pipeline. We used Molecfit \citep{Smette2015Molecfit:Correction,Kausch2015Molecfit:Correction} to correct for telluric contamination as the region around the Na\,\textsc{i} doublet is highly affected by the Earth's atmosphere. Barycentric radial velocity corrections were also applied to the spectra.

\subsection{UVES results}\label{sec:UVESresults}

Using data from the UVES spectrograph, we also searched for gas absorption towards HD\,44892, particularly in the Ca\,\textsc{ii} K \& H, and Na\,\textsc{i} D1 \& D2 lines. By fitting Kurucz model spectra to the stellar spectrum of our target, we identified a narrow absorption feature, located at 22.5\,km\,s$^{-1}$ (36\,$\sigma$, Figure\,\ref{fig:AbsAndSkyMap}). Although this feature does not correlate with the radial velocity of any known gas clouds in the line of sight and falls within the stellar radial velocity range of 20.88\,$\pm$\,1.85\,km\,s$^{-1}$, it is matched well by several of the nearby comparison stars. Interstellar origin of this feature is further supported by its strength in both of the sodium lines, which are a strong indication of interstellar medium (ISM) gas presence. The detailed description of methods used and analysis of the absorption features of this and other objects will be published in Szewczyk et al. (in prep).

There is, however, some variability in the residual flux present in both Ca\,\textsc{ii} K and H lines (Figure\,\ref{fig:AbsAndSkyMap}), which is not present in the sodium lines. This variability does not match with any of the clouds in the line of sight nor the nearby comparison stars. While not at the radial velocity of the parent star, this variability in the residual absorption might be caused by secondary gas released from volatile bodies like in case of $\beta$\,Pic \citep{Dent2014MolecularDisk} or a gas wind-disc, like the one found around NO\,Lup \citep{Lovell2021RapidDisc}, although this cannot be confirmed with the data we currently have. Thus, it is possible we have detected variable, circumstellar gas around HD\,44892.

Following the methods described in \citet{Fairlamb2017ALines} we calculated the mass accretion rate of HD\,44892 from the H$\alpha$ spectrum. We found accretion rates of $\log\dot{M}_{acc}$\,=\,-7.64$^{+0.11}_{-0.07}$$\,M_{\odot}\,yr^{-1}$ and $\log\dot{M}_{acc}$\,=\,-7.69$^{+0.11}_{-0.07}$\,$M_{\odot}\,yr^{-1}$ for the two observed epochs, which are consistent with those of other Herbig Ae/Be stars \citep{Wichittanakom2020TheStars}. This further suggests that HD\,44892 is a pre-MS star, with an estimated age of 2.1\,Myr, as accretion is characteristic of young stellar objects. Variability is also observed in the H$\alpha$ line (Szewczyk et al., in prep), similar to that seen in the Ca\,\textsc{ii} K \& H lines. Hence, the derived accretion rate may correspond to a temporary phase of high accretion and should therefore be regarded as an approximate estimate.

\section{Analysis \& discussion}

\subsection{ALMA} \label{sec:Analysis}

\subsubsection{Continuum}

The continuum emission we have detected around HD\,44892 is only marginally resolved along the major axis. Because of this, we are unable to estimate the orbital distance to the debris disc using our current data, and hence cannot determine the dust temperature, on which dust mass depends. To provide a more reliable estimate of the dust mass, we have utilised the `simalma' task, which is part of \textsc{CASA} package. We first created models, each containing a circular dust ring at various disc inclinations, located at a range of orbital distances from the star (with disc radii extending to 20-100\,au along the major axis), which we then input into `simalma'. Because the continuum emission is only marginally resolved, we did not consider the ring width as a free parameter and fixed it to 1\,px (0.2\,au), small enough to not influence the fit. The aim of our approach was to determine the dust ring size that would best mimic our observations - effectively, the maximum radius at which the dust can be placed without the belt becoming resolved. The simulations were carried out using the exact same observational setup as our real continuum observations - including observing time, precipitable water vapour (pwv), frequency, antenna setup, and other parameters - in order to accurately replicate the noise, resolution, and observed flux.

We used the \textsc{CASA} task `imfit' to compare the simulated images to the actual data, and found that a ring located approximately 40-45\,au from the star best reproduced the observed emission. Since the minor axis is not resolved, we cannot give an estimate to its size. Based on this distance, assuming the dust particles behave as black bodies, we estimated the dust temperature using Equation\,\ref{eq:Distance}:

\begin{equation}\label{eq:Distance}
    R=\sqrt{\frac{L_{\odot}}{16\pi\sigma T^{4}}}
\end{equation}

where R is the distance from the parent star to the debris belt, $L_{\odot}$ is the stellar luminosity, $\sigma$ is the Stefan-Boltzmann constant, and $T$ is the dust belt temperature. A debris belt distance of 40-45\,au corresponds to a dust temperature of $\sim$115\,K.

We calculated the minimum dust mass of HD\,44892 assuming optically thin, thermal emission, with Equation\,\ref{eq:DustMass}:

\begin{equation} \label{eq:DustMass}
M_{dust} = \frac{F_{\nu} \times D^{2}}{B_{\nu}(T) \times \kappa_{\nu}}
\end{equation}

where $F_{\nu}$ is the source emission, $D$ is the distance to the source, $B_{\nu}(T)$ is the Planck function (dependent on dust temperature, $T$), and $\kappa_{\nu}$ is the dust opacity, which is very uncertain. Here, we assume $\kappa=0.1\times2.3\times(\frac{\nu}{230\times10^9})^{0.7}$ [cm$^{2}$\,g$^{-1}$] \citep{Beckwith1990}, where $\nu$ is the frequency of observations [Hz]. This results in a minimum dust mass at the dust temperature of 115\,K of 0.019\,$\pm$\,0.009\,M$_{\oplus}$. This mass is highly dependent on the dust temperature, which is uncertain given the emission is only marginally resolved along the major axis. For comparison, a lower dust temperature of 50\,K would yield a higher dust mass estimate of 0.05\,$\pm$\,0.03$M_{\oplus}$.

Based on the expected lifetime of the observed dust, we can estimate whether it is primordial (left over from system formation) or secondary (released in planetesimal collisions). In the absence of gas, the lifetime of millimetre dust would be dominated by grain-grain collisions. The longest timescale at which this amount of millimetre dust can be destroyed, assuming collisions with dust of the same size, is if we distribute it in the entire region up to 40-45\,au. In this scenario, following \citet{Pearce2021Fomalhaut} \citep[Equation 8, based on][]{Wyatt2002Collisions} we obtain the dust lifetime of order $10^{-2}$\,Myr, which is much smaller than the estimated age of HD\,44892. Whilst the derived lifetime could be extended in the presence of gas, which can damp grains and reduce their impact velocities, very significant damping would be required to extend the lifetime up the the ${2\text{\,Myr}}$ star age. Hence the estimated lifetime, combined with the relatively low dust mass compared to protoplanetary discs (\mbox{Figure\,\ref{fig:DustAgeEvolution}}), leads us to believe that at least some of the dust is secondary in nature. This is typical for hybrid discs, which are thought to comprise primordial gas and second-generation dust (e.g. \citealt{Kospal2013ALMA21997, Moor2013ALMA21997}).

\subsubsection{Gas}

To calculate the CO gas mass in this disc we follow methods described in \citet{Miley2019Asymmetric100546} and \citet{Matra2015COALMA}. Since our detection is in ${}^{12}$CO (2-1) only, we use Equation\,\ref{eq:GasMass} to calculate the minimum CO gas mass, assuming local thermodynamic equilibrium:

\begin{equation} \label{eq:GasMass}
M_{CO}=\frac{4\pi m d^{2}}{h\nu_{21}A_{21}}\frac{F_{21}}{x_{2}}
\end{equation}

where $F_{21}$ is the integrated flux of gas emission, $d$ is the distance to the star, $m$ is the mass of the emitting molecule (${}^{12}$CO), $\nu_{21}$ is the rest frequency of the transition, $h$ is Planck's constant, $A_{21}$ is the appropriate Einstein coefficient, and $x_{2}$ is the fraction of the population found in the upper level of the transition (in this case `2'), estimated using a Boltzmann distribution. Since the gas emission is spatially unresolved, we cannot estimate its temperature. Using the flux measured for $^{12}$CO emission, for 20\,K gas we obtain a minimum CO gas mass of (7.86\,$\pm$\,2.05)\,$\times$\,10$^{-5}$\,$M_{\oplus}$ or (1.62\,$\pm$\,0.17)\,$\times$\,10$^{-4}$\,$M_{\oplus}$ assuming warmer gas at 50\,K. 

Given the young age of HD\,44892 and its transitional stage from a protoplanetary to a debris disc, the primordial origin of gas cannot be ruled out \citep{Nakatani2021PhotoevaporationRemnants}. Primordial gas is leftover from the protoplanetary stage, where it is dominated by the ISM-like abundances of molecular hydrogen (H$_{2}$), with carbon monoxide (CO) along other gases contributing to the total gas mass. Assuming the disc is hybrid, we use the ISM ratio of ${}^{12}\text{CO}$/H$_{2}$=$10^{-4}$ to estimate the total gas mass in this disc, under the assumption of a substantial hydrogen presence in the disc. If the gas is indeed primordial, we estimate the H$_{2}$ mass, and hence total disc gas mass, to be 0.06\,$\pm$\,0.02\,$M_{\oplus}$ for 20\,K gas and 0.12\,$\pm$\,0.01\,$M_{\oplus}$ for 50\,K gas. These estimates are based on the lower limit of CO gas mass (estimated from ${}^{12}\text{CO}$ emission), hence the total gas mass in this disc may be higher.

From the moment 0 map made for the two non-detections, ${}^{13}$CO and C$^{18}$O, we derived upper limits to the CO gas mass. By overlaying the 3\,$\sigma$ contours of ${}^{12}$CO emission on this map, we derived the RMS within this region to be 30.28\,mJy\,km\,s$^{-1}$, with 3$\sigma=90.84$\,mJy\,km\,s$^{-1}$, which we use as $^{13}$CO and C$^{18}$O flux for the purpose of gas mass calculations, since the dataset is common between them. We adopt the ISM-like relative ratios of CO/$^{13}$CO and CO/C$^{18}$O as 77 and 550, respectively \citep{Wilson1994AbundancesMedium}. Using Equation\,\ref{eq:GasMass} with adjusted values for $m$ and $x_{2}$, we obtained the upper limits to the CO gas mass as presented in Table\,\ref{tab:UpperLimits}. However, since in debris discs $^{13}$CO is typically more abundant than C$^{18}$O \citep[see, e.g.,][]{Miley2018Unlocking141569}, the upper limit on the CO gas mass derived from $^{13}$CO is more stringent. The total CO gas mass in the debris disc of HD\,44892 should hence be contained between the lower limits from the $^{12}$CO emission and upper limits from $^{13}$CO non-detection.

\begin{table}
    \centering
    \begin{tabular}{llll} \hline
        In units of\,$M_{\oplus}$: & $^{13}$CO & C$^{18}$O & $^{12}$CO \\ \hline
        $M_{CO}$ (20K) & <7.09\,$\times$\,10$^{-3}$ & <5.27\,$\times$\,10$^{-2}$ & (7.86\,$\pm$\,2.05)\,$\times$\,10$^{-5}$ \\ \hline
        $M_{CO}$ (50K) & <1.47\,$\times$\,10$^{-2}$  & <1.10\,$\times$\,10$^{-1}$ & (1.62\,$\pm$\,0.17)\,$\times$\,10$^{-4}$ \\ \hline
        $M_{H_{2}}$ (20K) & <5.06 & <37.67 & 0.06\,$\pm$\,0.02 \\ \hline
        $M_{H_{2}}$ (50K) & <10.51 & <78.30 & 0.12\,$\pm$\,0.01 \\ \hline
    \end{tabular}
    \caption{Upper limits to CO and total gas mass derived from $^{13}$CO and C$^{18}$O non-detections, and lower limits from $^{12}$CO emission.}
    \label{tab:UpperLimits}
\end{table}

\subsubsection{Gas and dust emission morphology}

\begin{figure}
    \centering
    \includegraphics[width=\columnwidth]{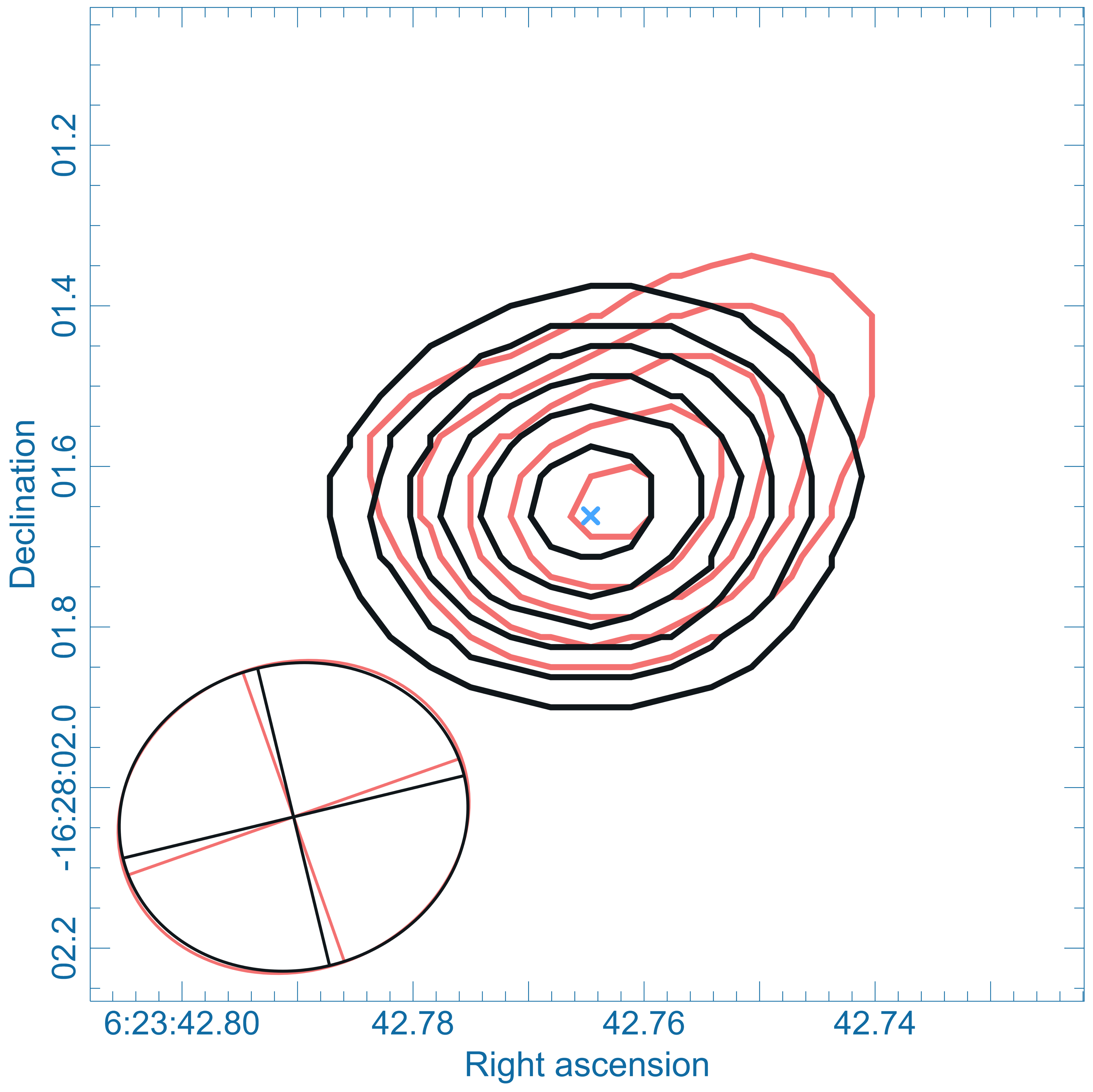}
    \caption{Overlaid contours of continuum (red) and CO emission (black) of HD\,44892. CO emission marked in contours of (3, 4, 5, 6, 7, 8)\,$\times$\,$\sigma$, where $\sigma=11.01$\,mJy\,beam$^{-1}$\,km\,s$^{-1}$, while continuum is (3, 4, 5, 6, 7)\,$\times$\,$\sigma$, with $\sigma=0.016$\,mJy\,beam$^{-1}$. Position of HD\,44892 marked as a blue cross.}
    \label{fig:HD44892gasCont}
\end{figure}

Examining the structure of both the continuum and $^{12}$CO emission (Figure\,\ref{fig:HD44892gasCont}), we can see that the dust emission clearly has a different morphology than the $^{12}$CO emission. This direct comparison in their morphology is possible because of the similar signal-to-noise ratio of both dust and $^{12}$CO emission. The $^{12}$CO integrated emission is not spatially resolved, therefore it follows the beam distribution. As mentioned in Section\,\ref{sec:dustFlux}, the continuum emission is marginally resolved, which can be further seen in Figure\,\ref{fig:ContMargResolved}. Here, the longest UV-distance baselines, at $\sim$800\,m, deviate from a flat profile, which indicates the continuum emission is not unresolved, since a point-like source would have a constant amplitude across all baselines. The extended continuum emission is only marginally resolved with an apparent asymmetry at a level of 4.5\,$\sigma$, which may or may not indicate asymmetry in the dust disc. The larger extent of continuum emission with respect to CO emission is atypical in protoplanetary discs, and more common in debris discs \citep[e.g.,][]{PPVII2023,Moor2019New32297,DiFolco2020}. Higher resolution observations are necessary to ascertain the distribution of dust and CO gas in the HD\,44892 system. 

\subsection{SED}

\begin{figure*}
    \centering
    \includegraphics[width=2\columnwidth]{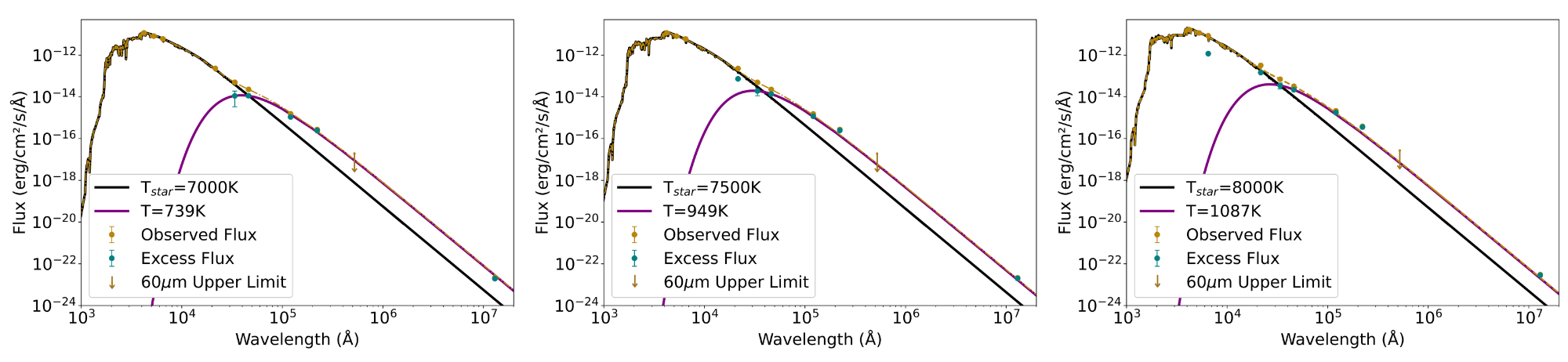}
    \caption{A fit to the SED of HD\,44892, for three assumed parent star temperatures: 7000\,K (left), 7500\,K (middle), and 8000\,K (right). These temperatures correspond to the uncertainty range reported by \citet{Iglesias2023X-shooterEvolution} (7500\,$\pm$\,500\,K). The photometry points used for all plots were V, B, \& R photometry, 2MASS ks, WISE W1-W4 \citep[all from][]{Iglesias2023X-shooterEvolution}, IRAS 60\,$\mu$m upper limit \citep[VOSA,][]{Bayo2008VOSA:Analyzer}, and ALMA continuum flux at 1.3\,mm as reported here in Section\,\ref{sec:dustFlux}, in order from shortest to longest wavelength. The fitted belt temperatures are shown for each case, using theoretical spectra from VOSA \citep{Bayo2008VOSA:Analyzer}.}
    \label{fig:SED}
\end{figure*}

Figure\,\ref{fig:SED} shows the SED of the HD\,44892 debris disc, assuming optically thin dust. The excess emission is best modelled with a single-temperature belt structure, fitted using a blackbody model. Due to the uncertainties in the temperature of the parent star T$_{eff}$\,=\,7500\,$\pm$\,500\,K, we present the SED fit to the stellar temperature of 7500\,K along with the alternative minimum and maximum values of 7000\,K and 8000\,K. The fit to the excess is highly dependent on the assumed stellar temperature, changing the resultant dust-belt temperatures. The best-fit SED (left plot of Figure\,\ref{fig:SED}, 7000\,K) suggests that the debris disc is composed of a warm belt at 739\,K, based on which we estimate the location of this component belt using Equation\,\ref{eq:Distance}.

We estimate the distance of the 739\,K dust belt to be 1.1\,au from the parent star. The two higher-temperature fits produce additional excess points that cannot be fitted with the existing blackbody component and would hence require an additional, even hotter belt fit. However, at such high temperatures, around A-type stars, it would be physically impossible for the dust to survive, suggesting the stellar fit might not be accurate. Carbon grains $\sim$0.1\,au away from A-type stars sublimate at $\sim$2000\,K \citep[e.g.,][]{Lebreton2013, Pearce2020}. Silicates sublimate at comparatively cooler temperatures of $\sim$1200\,K \citep[e.g.,][]{Sezestre2019}. In addition to the SED data shown in Figure\,\ref{fig:SED}, there exists a 100$\mu$m upper limit listed in the IRAS catalogue, which we did not include in the SED fit, as it is very large and does not represent a meaningful constraint.

\subsection{IRS}

\begin{figure}
    \centering
    \includegraphics[width=0.9\columnwidth]{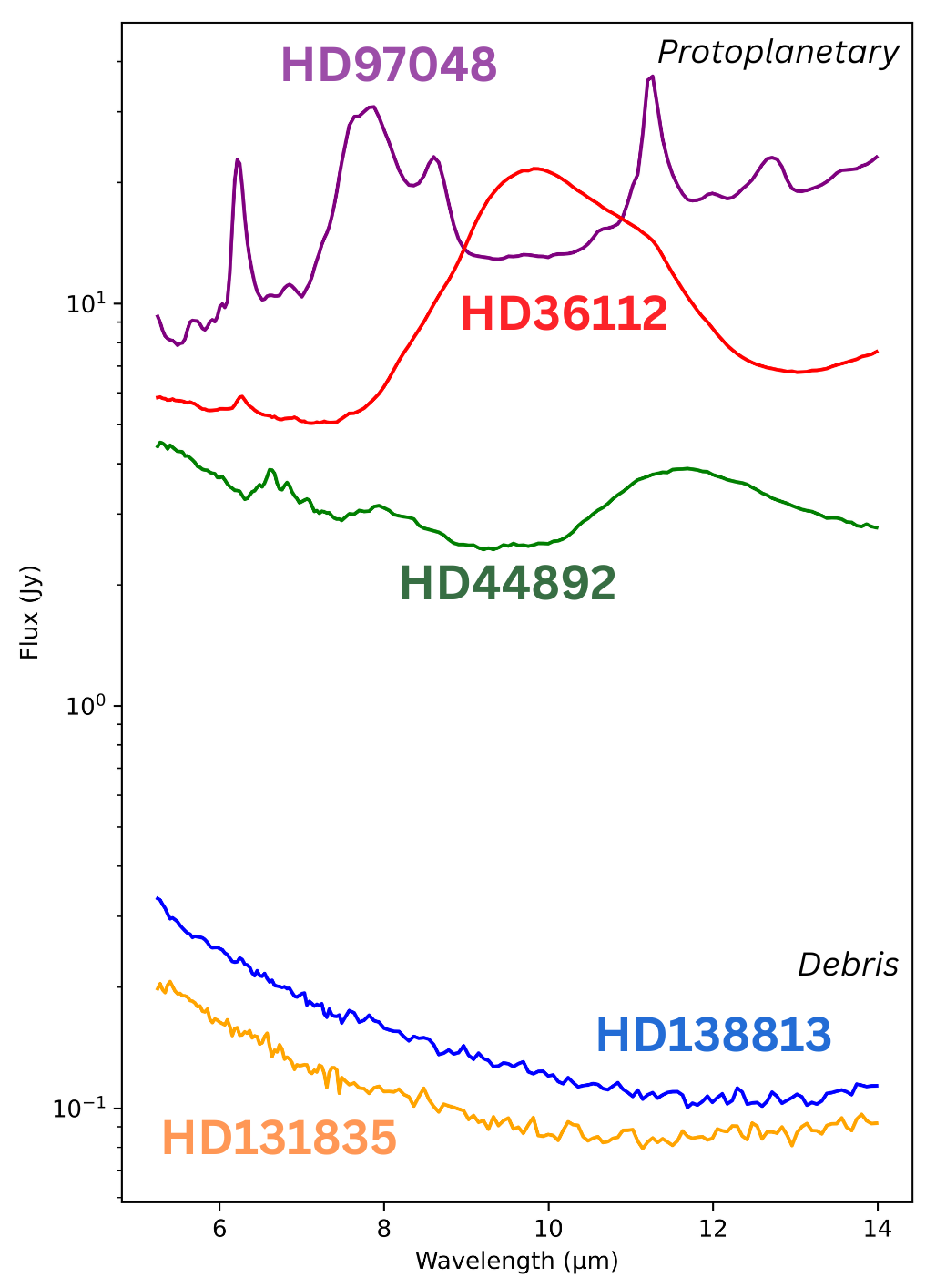}
    \caption{Comparison of mid-IR IRS spectra of HD\,44892 with those of protoplanetary discs (HD\,97048 and HD\,36112) and debris discs with gas (HD\,138813 and HD\,131835), all scaled to 100\,pc. HD\,44892 looks very different from both protoplanetary and debris discs, being caught in the transition between the two stages.}
    \label{fig:IRS}
\end{figure}

Using data from the IRS spectrograph\footnote{With reduced spectra obtained from \url{https://cassis.sirtf.com}} (the Infrared Spectrograph on the Spitzer Space Telescope), in Figure\,\ref{fig:IRS} we examined and compared the mid-IR spectrum of HD\,44892 (PI Bouwman, ID 3470) to that of protoplanetary discs HD\,97048 (PI Houck, ID 2) and HD\,36112 (PI Bouwman, ID 3470), and debris discs with gas HD\,138813 (PI Carpenter, ID 30091) and HD\,131835 (PI Houck, ID 40651). The spectrum of HD\,44892 is quite complex and looks very different to those of both protoplanetary and debris discs, and it shows a distinctive feature at the wavelength of $\sim$\,12\,$\mu$m, which is not present in the spectra of other objects. The comparison shows that HD\,44892 is unique and likely caught at the transition from protoplanetary to debris disc stage. 

\section{HD~44892 in the context of known gas-harbouring debris discs} \label{sec:Comparison}

\begin{figure*}
    \centering
    \includegraphics[width=1.8\columnwidth]{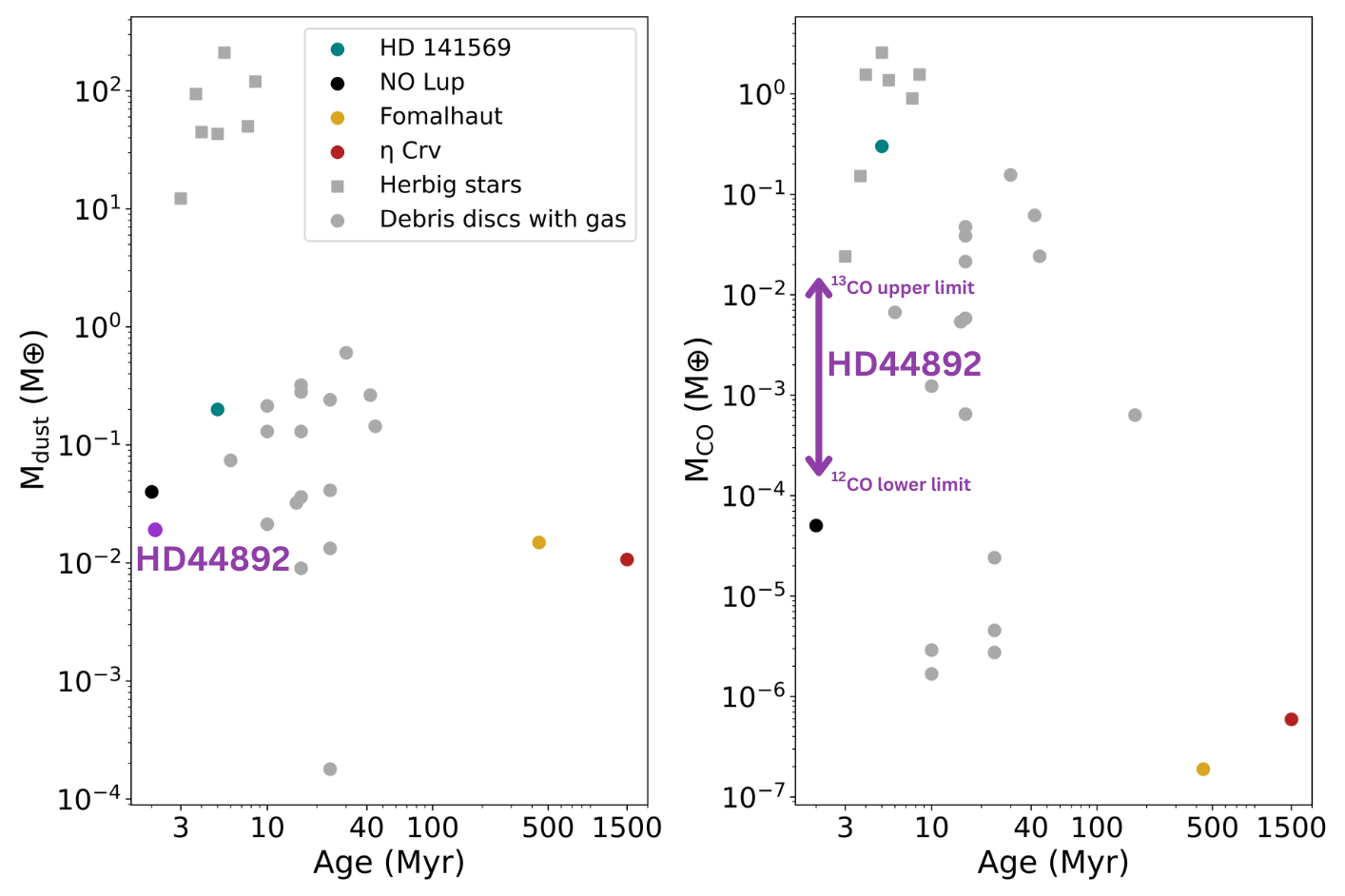}
    \caption{Comparison of the dust and CO masses of HD\,44892 (in purple) with other known debris discs hosting CO gas (grey points) and fainter Herbig star discs (grey squares, adapted from \citet{Moor2017MolecularStars}, calculated for $T=25K$). For reference we marked: the hybrid disc HD\,141569, with a dust mass of $0.20\pm0.07$$M_{\oplus}$ and a CO gas mass of $(3.01\pm0.48)\times10^{-1}$\,$M_{\odot}$ \citep{Miley2018Unlocking141569}; NO Lup, assumed to be 2\,Myr old for this plot, which has a CO gas mass of $(5.03\pm1.12\times10^{-5}$$M_{\oplus}$ and a dust mass of $(4.00\pm1.38)\times10^{-2}$$M_{\oplus}$ \citep{Lovell2021ALMADispersal}; Fomalhaut, at 440\,Myr, which has a CO gas mass of $(1.89\pm0.55)\times10^{-7}$$M_{\oplus}$ and a dust mass of $(1.49\pm0.38)\times10^{-2}$$M_{\oplus}$ \citep{Matra2017DetectionComets, MacGregor2017ADisk}; lastly, the 1-2\,Gyr-old system $\eta$ Crv, assumed to be 1.5\,Gyr for this plot, which has a CO gas mass of $(5.92\pm 0.78)\times10^{-7}$$M_{\oplus}$ and a dust mass of $(1.07\pm0.87)\times10^{-2}$$M_{\oplus}$ \citep{Marino2017ALMAPlanets}. All CO gas masses of debris discs with gas have been derived using a gas temperature of 50\,K, while for dust masses we used corresponding dust temperatures for each object. Remaining CO and dust fluxes used were adapted from from \citet{Moor2019New32297,Kospal2013ALMA21997, Moor2013ALMA21997,Rebollido2022The36546, Dent2014MolecularDisk,Matra2017ExocometaryObservations,Moor2017MolecularStars,Lieman-Sifry2016DebrisALMA,Miley2018Unlocking141569,Marino2016ExocometaryRing,Matra2019On7, Moor2019New32297, Bayo2019Sub-millimetreALMA, Su2017ALMASystem}; see Table\,\ref{tab:allCOGasDetections} for details.}
    \label{fig:MassComparison}
\end{figure*}

In Figure\,\ref{fig:MassComparison} we compare the dust and CO gas masses we calculated for HD\,44892 with Herbig star discs \citep[flux adapted from][]{Moor2017MolecularStars} and all other debris discs with CO gas mass measurements to date (see Table\,\ref{tab:allCOGasDetections} for detailed list), highlighting the discs HD\,141569 \citep{Miley2018Unlocking141569}, NO Lup \citep{Lovell2021ALMADispersal}, Fomalhaut \citep{Matra2017DetectionComets,MacGregor2017ADisk}, and $\eta$\,Crv \citep{Marino2017ALMAPlanets}. HD\,44892, due to its probable very young age, populates an area where we would expect protoplanetary discs to reside, but both its dust and CO gas masses are much lower than expected for PPDs. 

Since HD\,44892 is younger than almost all of debris discs from Figure\,\ref{fig:MassComparison}, we would expect it to contain more both dust and gas, as less time has passed for collisional evolution and subsequent dispersal of material from the system. While the CO gas mass of our target may be quite high and comparable to the most massive debris discs with gas, the dust mass is lower in comparison. Like HD\,141569, HD\,44892 is also a pre-main sequence star. Both these discs seem to represent the missing link between protoplanetary discs and the older planetary systems we observe around other stars. They are both at an evolutionary stage following PPDs, suggesting their gas is potentially inherited from this earlier stage, hence is primordial. This does not, however, entirely rule out the possibility of secondary or mixed origin gas.

The accretion rate we derived for HD\,44892 (see Section\,\ref{sec:UVESresults}) is very similar to that measured for HD\,141569 \citep{Fairlamb2015} and is orders of magnitude higher than expected for CO-rich debris discs \citep[10$^{-11}$-10$^{-13}$\,$M_{\odot}\,yr^{-1}$,][]{Kral2017PredictionsPlanetesimals,Borthakur2025}. If we assume a secondary origin for the gas, where the gas is H-poor with no significant H$_2$ reservoir, and adopting the upper limit on the total CO gas mass (based on the $^{13}$CO emission) of 10$^{-2}$\,$M_{\oplus}$, the CO in the disc would be depleted within just a few years. It is highly unlikely that a mechanism capable of replenishing such a large amount of secondary gas exists, thus the high accretion rate suggests that at least the inner disc is of primordial origin and dominated by H$_{2}$ gas. However, given the accretion variability discussed in Section\,\ref{sec:UVESresults}, this value may represent a phase of enhanced accretion. A more detailed discussion will be presented in Szewczyk et al. (in prep). 

At the young age of HD\,44892, apart from HD\,141569, only one other comparable $^{12}$CO detection in a debris disc exists - that in the disc around NO\,Lup, a 0.7\,$M_{\odot}$ star with an age estimate of 1-3\,Myr. The two other possible $^{12}$CO detections presented in \citet{Lovell2021ALMADispersal}, around Sz\,108B and Lup\,818s, are likely caused by CO cloud (ISM) gas instead. For NO\,Lup, using continuum and $^{12}$CO fluxes from \citet{Lovell2021ALMADispersal}, we derive the CO gas mass, assuming gas temperature of 50\,K, of (5.03\,$\pm$\,1.13)\,$\times$\,10$^{-5}$\,$M_{\oplus}$. At the reported dust temperature of 20\,K, we estimate the dust mass in the disc to be (4.00\,$\pm$\,1.38)\,$\times$\,10$^{-2}$\,$M_{\oplus}$. These values are comparable to those we derive for HD~44892. HD\,44892, similarly to NO\,Lup, has lost its protoplanetary disc, but retains some gas and dust. However, given their very different stellar masses and structure, the mechanisms responsible for dispersal are very likely different, i.e., FUV/EUV photoevaporation for the radiative A-type star HD~44892, and X-ray photoevaporation for the convective low mass star NO\,Lup \citep{Kunitomo2021PhotoevaporativeStars}.

The probable age of 2.1\,Myr makes HD\,44892 the youngest known gas-bearing debris disc around an IMS to date. There are several reasons why we trust this age determination the most. First, this object has a relatively high level of fractional IR excess (particularly at 12\,$\mu$m), putting it in the `hybrid' category. Such excess is observed to decrease with age, as shown in Figure\,\ref{fig:DustAgeEvolution}, and stars older than 100\,Myr are expected to have $F_{12\mu m}/F_{star}<3$. Given HD\,44892 has a fractional excess of 7.86 at this wavelength, it is unlikely to be very old, as no debris disc with such a high excess has been found at such an advanced age (see Figure\,\ref{fig:DustAgeEvolution}). 

Only three debris discs with gas have been found around stars older than 100\,Myr: HD\,23642 (V*\,V1229\,Tau) at 170\,Myr \citep{Pericaud2017TheDisks,Southworth2022ScutiHD23642}, Fomalhaut (HD\,216956) at 440\,Myr \citep{Matra2017DetectionComets}, and the 1-2\,Gyr $\eta$\,Crv \citep{Marino2017ALMAPlanets}. The older age possibility of HD\,44892, at 800\,Myr, would make it comparable to the latter two of these objects. For Fomalhaut, using data from \citet{Matra2017DetectionComets} and \citet{MacGregor2017ADisk}, we derive a 50\,K dust mass of $(1.49\pm0.38)\times10^{-2}$$M_{\oplus}$ and a CO gas mass of $(1.89\pm0.55)\times10^{-7}$$M_{\oplus}$. For $\eta$\,Crv we derive a dust mass at 50\,K of $(1.07\pm0.87)\times10^{-2}$$M_{\oplus}$ and a CO gas mass of $(5.92\pm0.78)\times10^{-7}$$M_{\oplus}$ \citep[with flux data from][]{Marino2017ALMAPlanets}. Both systems contain at least three orders of magnitude less CO gas mass than HD\,44892 (see Figure\,\ref{fig:MassComparison}). This further shows that if HD\,44892 was 800\,Myr, it would be an outlier, containing significantly high amounts of CO gas.

It is worth noting that discs exist around post-asymptotic giant branch (AGB) binary star systems, with dust masses even up to 1\,$M_{J}$ \citep{Kluska2022APlanets}. These, however, are created from the mass-loss of primary object during the last stages of its evolution. HD\,44892, at the age of 2.1\,Myr, is approaching the zero-age main sequence (ZAMS) or, at the alternative age of 800\,Myr, is crossing the Hertzsprung-Russell (HR) gap after leaving the MS. Both its stellar temperature and luminosity are inconsistent with the star being on the AGB. Moreover, there is no evidence to suggest that HD\,44892 is a binary system, hence this scenario is inapplicable here.

\section{Summary and future prospects}

Using ALMA, we have detected CO emission in a debris disc around a young, pre-MS intermediate-mass star HD\,44892 with a fractional IR excess of 7.86\,$^{+0.11}_{-2.27}$ at 12\,$\mu$m \citep{Iglesias2023X-shooterEvolution}, which falls within a range ($3<F_{12\mu m}/F_{star}<25$) where only one such object, HD\,141569, has been previously identified with a measured CO gas mass. This provides a rare glimpse into the very stage of transition from protoplanetary to debris discs. HD\,44892 is hence also one of the youngest known hybrid discs, along with the lower-mass (0.7\,$M_{\odot}$) star NO~Lup, with an age 1-3\,Myr \citep{Lovell2021ALMADispersal}. For gas temperature in the range 20-50~K, the CO detection yields an estimate of CO gas mass of around 10$^{-4}$\,$M_{\oplus}$. On the other hand, the non-detections of $^{13}$CO and C$^{18}$O lines provide upper limits on gas mass of around 10$^{-2}$\,$M_{\oplus}$. Since we do not know whether the $^{12}$CO emission we detected is optically thin or thick, the CO gas mass we derived from $^{12}$CO emission should be treated as a lower limit, and may be up to $\sim$10$^{-2}$\,$M_{\oplus}$, which is the upper limit derived from $^{13}$CO non-detection.

Given the possibility of the primordial origin of gas, we also report the total gas mass values, calculated assuming an ISM ratio of ${}^{12}\text{CO}$$/$H$_{2}$=$10^{-4}$ in the disc. This results in total H$_2$ gas mass of 0.06\,$\pm$\,0.02\,$M_{\oplus}$ for 20\,K gas and 0.12\,$\pm$\,0.01\,$M_{\oplus}$ for 50\,K gas, yielding a gas-to-dust mass ratio of $\sim$3 at 20\,K or $\sim$6 at 50\,K. The H$_2$ masses presented are based on $^{12}$CO emission, which may be optically thick, and hence represent lower limits to the total disc gas mass. However, if this gas is of secondary rather than primordial origin, this ratio does not apply, since secondary gas is hydrogen-depleted \citep{Smirnov-Pinchukov2022LackGas}. In such a case, the gas is most likely CO dominated, making the CO gas mass a better representative of the total disc gas mass. 

Presence of circumstellar gas is further confirmed in our UVES observations of gas absorption, for which we were able to rule out interstellar origin. From the UVES data we also derived an accretion rate for HD\,44892 of $\log\dot{M}_{acc}$\,=\,-7.64$^{+0.11}_{-0.07}$$\,M_{\odot}\,yr^{-1}$ and $\log\dot{M}_{acc}$\,=\,-7.69$^{+0.11}_{-0.07}$\,$M_{\odot}\,yr^{-1}$ for the two observed epochs. This value is comparable to that of HD\,141569 exceeding the level expected for CO-rich debris discs and is similar to values found in protoplanetary discs. If the gas was secondary, then based on the $\sim10^{-2}$\,M$_{\oplus}$ CO gas mass upper limit, from our $^{13}$CO non-detection, it would be depleted on very short timescales, and replenishment at the required rate seems unlikely. This strongly suggests that at least the inner disc retains primordial, $H_2$-rich gas.

We also detect 1.3\,mm continuum emission in HD\,44892 in our ALMA observations. We report the dust mass of HD\,44892, assuming a dust temperature of 115\,K, of 0.019\,$\pm$\,0.009\,M$_{\oplus}$. Millimetre dust in this young disc is likely to originate from collisional evolution, given the collisional dust lifetime we estimate at $\sim$10$^{-2}$\,\text{Myr}. With the detected 1.3\,mm flux and known shorter-wavelength photometric measurements of HD\,44892, along with the stellar spectrum models, we attempt to characterise the SED in the context of single-temperature ring models. This approach suggests that almost all of the excess emission comes from hot material, over 700~K, thus close to the star. Indeed, most of the millimetre flux in HD\,44892 is unresolved and may originate from a point source. 

If this is correct, then the marginally resolved dust and gas extending at $\sim$40-45~au are very cold and vertically thin enough not to intercept stellar light - the only scenario to explain the lack of significant infrared emission from these spatial scales. Such flattened disc structures (and low-excess SEDs) may arise at late stages of protoplanetary disc evolution due to gas loss but also when all the very small, sub-micron dust has been removed \citep{Panic2017EffectsLocation}. Important to note is that removal of such dust creates conditions for gas survival beyond protoplanetary disc phase suggested by \citet{Nakatani2021PhotoevaporationRemnants}. Assuming HD\,44892 is an example of this scenario, future studies could involve scattered-light observations using telescopes such as the VLT/SPHERE, to verify if such small grains are indeed absent from its debris disc by, for example, if scattered light is absent, or has a colour indicative of scattering by larger grains \citep{Mulders2013WhyHD100546}. 

The dust mass in the disc of HD\,44892 is within an order of magnitude less than that of the hybrid disc HD\,141569. The CO gas mass may be comparable to that of known gas-rich debris discs, depending on the optical thickness of $^{12}$CO emission. Given the very young age of HD\,44892, primordial origin gas is likely \citep{Nakatani2021PhotoevaporationRemnants,Smirnov-Pinchukov2022LackGas}. Hence, it is possible that we have caught this disc just at the end of its primordial gas contents, which may explain the relatively low CO mass. This suggests HD\,44892 is the next evolutionary stage following HD\,141569, helping us understand the evolution of planetary systems better. Nonetheless, the young age of this object alone does not allow to make definitive conclusions about the origin of gas in this disc. 

It is important to note that from the CO gas mass alone, of (1.62\,$\pm$\,0.17)\,$\times$\,10$^{-4}$$M_{\oplus}$, it is impossible to fully rule out the secondary origin scenario. According to \citet{Cataldi2023PrimordialALMA}, only CO-rich debris discs, which they classify as having M$_{CO}\gtrsim10^{-3}$$M_{\oplus}$, cannot be solely explained by the current secondary models \citep[like the one in][]{Kral2017PredictionsPlanetesimals}. A solution may be either an improvement of secondary models of gas production or presence of primordial gas. On the other hand, they state that it is the debris discs with low CO gas masses, $\lesssim10^{-5}$$M_{\oplus}$ that are easily explainable by secondary means alone. Since the CO gas mass of HD\,44892 is between the two, secondary or mixed origin is still possible. It is worth noting, however, that while models introduced in \citet{Kral2019ImagingDiscs} can explain the high CO gas masses in CO-rich debris discs, by introducing neutral carbon shielding and self-shielding of CO, more recent work of \citet{Marino2022VerticalLifetime} discusses the limitations of this scenario. In particular, they show that vertical mixing of gas in the debris disc reduces the effectiveness of shielding, limiting the accumulation of CO. Hence, for the most CO-rich systems secondary models alone are not yet able to explain this gas presence, and primordial origin remains the best available explanation \citep{Cataldi2023PrimordialALMA, Marino2022VerticalLifetime}. \citet{Cataldi2023PrimordialALMA} propose that observations of C\,\textsc{i} in debris discs where CO was found are needed to improve current models of gas, which might help answer questions regarding its origin. Since HD\,44892 has a CO gas mass comparable to some of the debris discs with detected C\,\textsc{i} in their sample, it is possible that HD\,44892 may also have detectable C\,\textsc{i} emission in higher sensitivity observations. 

If the alternative age of 800\,Myr is considered, HD\,44892 would be an outlier in terms of its dust mass (temperature-dependent), 12\,$\mu$m excess, and CO gas mass, strongly suggesting that this is the case of a young pre-main sequence star. The two proposed age scenarios, of 2.1\,Myr and 800\,Myr, correspond to different stellar mass estimates of 2.87\,$M_{\odot}$ and 2.29\,$M_{\odot}$, respectively. Future, higher spatial and spectral resolution studies of CO emission with ALMA, could remove any remaining uncertainties, allowing for a dynamical mass estimation of the star. 

\section*{Acknowledgements}

KS acknowledges funding from the Bell Burnell Graduate Scholarship Fund (BB0023). OP and DPI acknowledge support from the Science and Technology Facilities Council via grant number ST/X001016/1. TDP is supported by a UKRI Stephen Hawking Fellowship and a Warwick Prize Fellowship, the latter made possible by a generous philanthropic donation. JM acknowledges support from FONDECYT de Postdoctorado 2024 \#3240612 and ANID -- Millennium Science Initiative Program -- Center Code NCN2024\_001.

We thank Lionell Siess, Sebastian Marino, Rens Waters, and Till Käufer for fruitful discussions. 

We thank the anonymous referee for their insightful comments and suggestions, which provided valuable perspectives on the nature of this system.

This paper makes use of the following ALMA data: ADS\allowbreak/JAO.ALMA\#2022.1.01686.S 
ALMA is a partnership of ESO (representing its member states), NSF (USA), and NINS (Japan).  
It includes NRC (Canada), MOST and ASIAA (Taiwan), and KASI (Republic of Korea), in cooperation with the Republic of Chile.  
The Joint ALMA Observatory is operated by ESO, AUI/NRAO, and NAOJ.

Based on observations collected at the European Southern Observatory under ESO programmes 112.262.P.001, 094.A-9029(D), 082.D-0499(A), and 085.D-0395(A).

This work has made use of data from the European Space Agency (ESA) mission
{\it Gaia} (\url{https://www.cosmos.esa.int/gaia}), processed by the {\it Gaia}
Data Processing and Analysis Consortium (DPAC,
\url{https://www.cosmos.esa.int/web/gaia/dpac/consortium}). Funding for the DPAC
has been provided by national institutions, in particular the institutions
participating in the {\it Gaia} Multilateral Agreement.

This publication makes use of VOSA, developed under the Spanish Virtual Observatory (\url{https://svo.cab.inta-csic.es}) project funded by MCIN/AEI/10.13039/501100011033/ through grant PID2020-112949GB-I00.
VOSA has been partially updated by using funding from the European Union's Horizon 2020 Research and Innovation Programme, under Grant Agreement n$\degree$ 776403 (EXOPLANETS-A).

This work is based [in part] on observations made with the Spitzer Space Telescope, which was operated by the Jet Propulsion Laboratory, California Institute of Technology under a contract with NASA.

%%%%%%%%%%%%%%%%%%%%%%%%%%%%%%%%%%%%%%%%%%%%%%%%%%
\section*{Data Availability}

All data underlying this article are available at the ALMA, ESO, and GAIA archives. ALMA data is found under project 2022.1.01686.S at \url{https://almascience.eso.org/aq/}. ESO data
are found under projects 112.262.P.001, 094.A-9029(D), 082.D-0499(A), and 085.D-0395(A) at
\url{http://archive.eso.org/cms.html}. GAIA archive can be browsed at
\url{https://gea.esac.esa.int/archive}. Spitzer IRS spectra can be found at \url{https://cassis.sirtf.com}.



\bibliographystyle{mnras}
\bibliography{example} % if your bibtex file is called example.bib

@article{Abt1973RotationDwarfs.,
	  author = {{Abt}, H.~A. and {Moyd}, K.~I.},
        title = "{Rotation and shell spectra among A-type dwarfs.}",
      journal = {\apj},
         year = 1973,
        month = jun,
       volume = {182},
        pages = {809},
          doi = {10.1086/152184},
       adsurl = {https://ui.adsabs.harvard.edu/abs/1973ApJ...182..809A}
}

@article{AllendePrieto1999FundamentalTemperatures,
	author = {{Allende Prieto}, C. and {Lambert}, D.~L.},
        title = "{Fundamental parameters of nearby stars from the comparison with evolutionary calculations: masses, radii and effective temperatures}",
      journal = {\aap},
         year = 1999,
        month = dec,
       volume = {352},
        pages = {555-562},
          doi = {10.48550/arXiv.astro-ph/9911002},
archivePrefix = {arXiv},
       eprint = {astro-ph/9911002},
 primaryClass = {astro-ph},
       adsurl = {https://ui.adsabs.harvard.edu/abs/1999A&A...352..555A}
}

@article{Ballester2000ThePipeline,
	 author = {{Ballester}, P. and {Modigliani}, A. and {Boitquin}, O. and {Cristiani}, S. and {Hanuschik}, R. and {Kaufer}, A. and {Wolf}, S.},
        title = "{The UVES Data Reduction Pipeline}",
      journal = {The Messenger},
         year = 2000,
        month = sep,
       volume = {101},
        pages = {31-36},
       adsurl = {https://ui.adsabs.harvard.edu/abs/2000Msngr.101...31B}
}

@article{Bayo2008VOSA:Analyzer,
	author = {{Bayo}, A. and {Rodrigo}, C. and {Barrado Y Navascu{\'e}s}, D. and {Solano}, E. and {Guti{\'e}rrez}, R. and {Morales-Calder{\'o}n}, M. and {Allard}, F.},
        title = "{VOSA: virtual observatory SED analyzer. An application to the Collinder 69 open cluster}",
      journal = {\aap},
         year = 2008,
        month = dec,
       volume = {492},
       number = {1},
        pages = {277-287},
          doi = {10.1051/0004-6361:200810395},
archivePrefix = {arXiv},
       eprint = {0808.0270},
 primaryClass = {astro-ph},
       adsurl = {https://ui.adsabs.harvard.edu/abs/2008A&A...492..277B}
}

@article{Bochanski2018FundamentalGaia,
	 author = {{Bochanski}, John J. and {Faherty}, Jacqueline K. and {Gagn{\'e}}, Jonathan and {Nelson}, Olivia and {Coker}, Kristina and {Smithka}, Iliya and {Desir}, Deion and {Vasquez}, Chelsea},
        title = "{Fundamental Properties of Co-moving Stars Observed by Gaia}",
      journal = {\aj},
         year = 2018,
        month = apr,
       volume = {155},
       number = {4},
          eid = {149},
        pages = {149},
          doi = {10.3847/1538-3881/aaaebe},
archivePrefix = {arXiv},
       eprint = {1801.00537},
 primaryClass = {astro-ph.SR},
       adsurl = {https://ui.adsabs.harvard.edu/abs/2018AJ....155..149B}
}

@article{Bayo2019Sub-millimetreALMA,
	author = {{Bayo}, A. and {Olofsson}, J. and {Matr{\`a}}, L. and {Beam{\'\i}n}, J.~C. and {Gallardo}, J. and {de Gregorio-Monsalvo}, I. and {Booth}, M. and {Zamora}, C. and {Iglesias}, D. and {Henning}, Th and {Schreiber}, M.~R. and {C{\'a}ceres}, C.},
        title = "{Sub-millimetre non-contaminated detection of the disc around TWA 7 by ALMA}",
      journal = {\mnras},
         year = 2019,
        month = jul,
       volume = {486},
       number = {4},
        pages = {5552-5557},
          doi = {10.1093/mnras/stz1133},
archivePrefix = {arXiv},
       eprint = {1806.09252},
 primaryClass = {astro-ph.SR},
       adsurl = {https://ui.adsabs.harvard.edu/abs/2019MNRAS.486.5552B}
}

@article{Cataldi2023PrimordialALMA,
	author = {{Cataldi}, Gianni and {Aikawa}, Yuri and {Iwasaki}, Kazunari and {Marino}, Sebastian and {Brandeker}, Alexis and {Hales}, Antonio and {Henning}, Thomas and {Higuchi}, Aya E. and {Hughes}, A. Meredith and {Janson}, Markus and {Kral}, Quentin and {Matr{\`a}}, Luca and {Mo{\'o}r}, Attila and {Olofsson}, G{\"o}ran and {Redfield}, Seth and {Roberge}, Aki},
        title = "{Primordial or Secondary? Testing Models of Debris Disk Gas with ALMA}",
      journal = {\apj},
         year = 2023,
        month = jul,
       volume = {951},
       number = {2},
          eid = {111},
        pages = {111},
          doi = {10.3847/1538-4357/acd6f3},
archivePrefix = {arXiv},
       eprint = {2305.12093},
 primaryClass = {astro-ph.EP},
       adsurl = {https://ui.adsabs.harvard.edu/abs/2023ApJ...951..111C}
}

@article{Chen2003TheHerculis,
	author = {{Chen}, C.~H. and {Jura}, M.},
        title = "{The Low-Velocity Wind from the Circumstellar Matter around the B9 V Star {\ensuremath{\sigma}} Herculis}",
      journal = {\apj},
         year = 2003,
        month = jan,
       volume = {582},
       number = {1},
        pages = {443-448},
          doi = {10.1086/344589},
archivePrefix = {arXiv},
       eprint = {astro-ph/0209076},
 primaryClass = {astro-ph},
       adsurl = {https://ui.adsabs.harvard.edu/abs/2003ApJ...582..443C}
}

@misc{Comrie2018CARTA:Astronomy,
  author       = {Angus Comrie and
                  Kuo-Song Wang and
                  Shou-Chieh Hsu and
                  Anthony Moraghan and
                  Pamela Harris and
                  Qi Pang and
                  Adrianna Pińska and
                  Cheng-Chin Chiang and
                  Tien-Hao Chang and
                  Yu-Hsuan Hwang and
                  Hengtai Jan and
                  Ming-Yi Lin and
                  Rob Simmonds},
  title        = {CARTA: The Cube Analysis and Rendering Tool for
                   Astronomy
                  },
  month        = jun,
  year         = 2021,
  publisher    = {Zenodo},
  version      = {2.0.0},
}

@INPROCEEDINGS{Dekker2000DesignObservatory,
       author = {{Dekker}, Hans and {D'Odorico}, Sandro and {Kaufer}, Andreas and {Delabre}, Bernard and {Kotzlowski}, Heinz},
        title = "{Design, construction, and performance of UVES, the echelle spectrograph for the UT2 Kueyen Telescope at the ESO Paranal Observatory}",
    booktitle = {Optical and IR Telescope Instrumentation and Detectors},
         year = 2000,
       editor = {{Iye}, Masanori and {Moorwood}, Alan F.},
       series = {Society of Photo-Optical Instrumentation Engineers (SPIE) Conference Series},
       volume = {4008},
        month = aug,
        pages = {534-545},
}

@article{Dent2014MolecularDisk,
	 author = {{Dent}, W.~R.~F. and {Wyatt}, M.~C. and {Roberge}, A. and {Augereau}, J. -C. and {Casassus}, S. and {Corder}, S. and {Greaves}, J.~S. and {de Gregorio-Monsalvo}, I. and {Hales}, A. and {Jackson}, A.~P. and {Hughes}, A. Meredith and {Lagrange}, A. -M. and {Matthews}, B. and {Wilner}, D.},
        title = "{Molecular Gas Clumps from the Destruction of Icy Bodies in the {\ensuremath{\beta}} Pictoris Debris Disk}",
      journal = {Science},
         year = 2014,
        month = mar,
       volume = {343},
       number = {6178},
        pages = {1490-1492},
          doi = {10.1126/science.1248726},
archivePrefix = {arXiv},
       eprint = {1404.1380},
 primaryClass = {astro-ph.SR},
       adsurl = {https://ui.adsabs.harvard.edu/abs/2014Sci...343.1490D},
}

@article{Esposito2020DebrisCampaign,
	author = {{Esposito}, Thomas M. and {Kalas}, Paul and {Fitzgerald}, Michael P. and {Millar-Blanchaer}, Maxwell A. and {Duch{\^e}ne}, Gaspard and {Patience}, Jennifer and {Hom}, Justin and {Perrin}, Marshall D. and {De Rosa}, Robert J. and {Chiang}, Eugene and {Czekala}, Ian and {Macintosh}, Bruce and {Graham}, James R. and {Ansdell}, Megan and {Arriaga}, Pauline and {Bruzzone}, Sebastian and {Bulger}, Joanna and {Chen}, Christine H. and {Cotten}, Tara and {Dong}, Ruobing and {Draper}, Zachary H. and {Follette}, Katherine B. and {Hung}, Li-Wei and {Lopez}, Ronald and {Matthews}, Brenda C. and {Mazoyer}, Johan and {Metchev}, Stan and {Rameau}, Julien and {Ren}, Bin and {Rice}, Malena and {Song}, Inseok and {Stahl}, Kevin and {Wang}, Jason and {Wolff}, Schuyler and {Zuckerman}, Ben and {Ammons}, S. Mark and {Bailey}, Vanessa P. and {Barman}, Travis and {Chilcote}, Jeffrey and {Doyon}, Rene and {Gerard}, Benjamin L. and {Goodsell}, Stephen J. and {Greenbaum}, Alexandra Z. and {Hibon}, Pascale and {Hinkley}, Sasha and {Ingraham}, Patrick and {Konopacky}, Quinn and {Maire}, J{\'e}r{\^o}me and {Marchis}, Franck and {Marley}, Mark S. and {Marois}, Christian and {Nielsen}, Eric L. and {Oppenheimer}, Rebecca and {Palmer}, David and {Poyneer}, Lisa and {Pueyo}, Laurent and {Rajan}, Abhijith and {Rantakyr{\"o}}, Fredrik T. and {Ruffio}, Jean-Baptiste and {Savransky}, Dmitry and {Schneider}, Adam C. and {Sivaramakrishnan}, Anand and {Soummer}, R{\'e}mi and {Thomas}, Sandrine and {Ward-Duong}, Kimberly},
        title = "{Debris Disk Results from the Gemini Planet Imager Exoplanet Survey's Polarimetric Imaging Campaign}",
      journal = {\aj},
         year = 2020,
        month = jul,
       volume = {160},
       number = {1},
          eid = {24},
        pages = {24},
          doi = {10.3847/1538-3881/ab9199},
archivePrefix = {arXiv},
       eprint = {2004.13722},
 primaryClass = {astro-ph.EP},
       adsurl = {https://ui.adsabs.harvard.edu/abs/2020AJ....160...24E}
}

@article{Fairlamb2017ALines,
	author = {{Fairlamb}, J.~R. and {Oudmaijer}, R.~D. and {Mendigutia}, I. and {Ilee}, J.~D. and {van den Ancker}, M.~E.},
        title = "{A spectroscopic survey of Herbig Ae/Be stars with X-Shooter - II. Accretion diagnostic lines}",
      journal = {\mnras},
         year = 2017,
        month = feb,
       volume = {464},
       number = {4},
        pages = {4721-4735},
          doi = {10.1093/mnras/stw2643},
archivePrefix = {arXiv},
       eprint = {1610.09636},
 primaryClass = {astro-ph.SR},
       adsurl = {https://ui.adsabs.harvard.edu/abs/2017MNRAS.464.4721F}
}

@article{Freudling2013AutomatedAstronomy,
	author = {{Freudling}, W. and {Romaniello}, M. and {Bramich}, D.~M. and {Ballester}, P. and {Forchi}, V. and {Garc{\'\i}a-Dabl{\'o}}, C.~E. and {Moehler}, S. and {Neeser}, M.~J.},
        title = "{Automated data reduction workflows for astronomy. The ESO Reflex environment}",
      journal = {\aap},
         year = 2013,
        month = nov,
       volume = {559},
          eid = {A96},
        pages = {A96},
          doi = {10.1051/0004-6361/201322494},
archivePrefix = {arXiv},
       eprint = {1311.5411},
 primaryClass = {astro-ph.IM},
       adsurl = {https://ui.adsabs.harvard.edu/abs/2013A&A...559A..96F}
}

@article{Gagne2018BANYAN.Pc,
	author = {{Gagn{\'e}}, Jonathan and {Mamajek}, Eric E. and {Malo}, Lison and {Riedel}, Adric and {Rodriguez}, David and {Lafreni{\`e}re}, David and {Faherty}, Jacqueline K. and {Roy-Loubier}, Olivier and {Pueyo}, Laurent and {Robin}, Annie C. and {Doyon}, Ren{\'e}},
        title = "{BANYAN. XI. The BANYAN {\ensuremath{\Sigma}} Multivariate Bayesian Algorithm to Identify Members of Young Associations with 150 pc}",
      journal = {\apj},
         year = 2018,
        month = mar,
       volume = {856},
       number = {1},
          eid = {23},
        pages = {23},
          doi = {10.3847/1538-4357/aaae09},
archivePrefix = {arXiv},
       eprint = {1801.09051},
 primaryClass = {astro-ph.SR},
       adsurl = {https://ui.adsabs.harvard.edu/abs/2018ApJ...856...23G}
}

@article{Gontcharov2006PulkovoSystem,
	 author = {{Gontcharov}, G.~A.},
        title = "{Pulkovo Compilation of Radial Velocities for 35 495 Hipparcos stars in a common system}",
      journal = {Astronomy Letters},
         year = 2006,
        month = nov,
       volume = {32},
       number = {11},
        pages = {759-771},
          doi = {10.1134/S1063773706110065},
archivePrefix = {arXiv},
       eprint = {1606.08053},
 primaryClass = {astro-ph.SR},
       adsurl = {https://ui.adsabs.harvard.edu/abs/2006AstL...32..759G}
}

@article{Hales2022ALMADisk,
	 author = {{Hales}, Antonio S. and {Marino}, Sebasti{\'a}n and {Sheehan}, Patrick D. and {Ulloa}, Silvio and {P{\'e}rez}, Sebasti{\'a}n and {Matr{\`a}}, Luca and {Kral}, Quentin and {Wyatt}, Mark and {Dent}, William and {Carpenter}, John},
        title = "{ALMA Observations of the HD 110058 Debris Disk}",
      journal = {\apj},
         year = 2022,
        month = dec,
       volume = {940},
       number = {2},
          eid = {161},
        pages = {161},
          doi = {10.3847/1538-4357/ac9cd3},
archivePrefix = {arXiv},
       eprint = {2210.12275},
 primaryClass = {astro-ph.SR},
       adsurl = {https://ui.adsabs.harvard.edu/abs/2022ApJ...940..161H}
}

@article{Hughes2018DebrisVariability,
	author = {{Hughes}, A. Meredith and {Duch{\^e}ne}, Gaspard and {Matthews}, Brenda C.},
        title = "{Debris Disks: Structure, Composition, and Variability}",
      journal = {\araa},
         year = 2018,
        month = sep,
       volume = {56},
        pages = {541-591},
          doi = {10.1146/annurev-astro-081817-052035},
archivePrefix = {arXiv},
       eprint = {1802.04313},
 primaryClass = {astro-ph.EP},
       adsurl = {https://ui.adsabs.harvard.edu/abs/2018ARA&A..56..541H}
}

@misc{Iglesias2020SearchingDisks,
  author       = {Iglesias, Daniela},
  title        = {Searching for gas in debris disks},
  month        = apr,
  year         = 2020,
  publisher    = {Zenodo},
  doi          = {10.5281/zenodo.14499893}
}

@article{Iglesias2018DebrisOrigin,
	author = {{Iglesias}, D. and {Bayo}, A. and {Olofsson}, J. and {Wahhaj}, Z. and {Eiroa}, C. and {Montesinos}, B. and {Rebollido}, I. and {Smoker}, J. and {Sbordone}, L. and {Schreiber}, M.~R. and {Henning}, Th},
        title = "{Debris discs with multiple absorption features in metallic lines: circumstellar or interstellar origin?}",
      journal = {\mnras},
         year = 2018,
        month = oct,
       volume = {480},
       number = {1},
        pages = {488-520},
          doi = {10.1093/mnras/sty1724},
archivePrefix = {arXiv},
       eprint = {1806.10687},
 primaryClass = {astro-ph.SR},
       adsurl = {https://ui.adsabs.harvard.edu/abs/2018MNRAS.480..488I}
}

@article{Iglesias2019AnDisc,
	author = {{Iglesias}, Daniela P. and {Olofsson}, Johan and {Bayo}, Amelia and {Zieba}, Sebastian and {Montesinos}, Mat{\'\i}as and {Smoker}, Jonathan and {Kennedy}, Grant M. and {Godoy}, Nicol{\'a}s and {Pantoja}, Blake and {Talens}, Geert Jan and {Wahhaj}, Zahed and {Zamora}, Catalina},
        title = "{An unusually large gaseous transit in a debris disc}",
      journal = {\mnras},
         year = 2019,
        month = dec,
       volume = {490},
       number = {4},
        pages = {5218-5227},
          doi = {10.1093/mnras/stz2888},
archivePrefix = {arXiv},
       eprint = {1910.04747},
 primaryClass = {astro-ph.EP},
       adsurl = {https://ui.adsabs.harvard.edu/abs/2019MNRAS.490.5218I}
}

@article{Iglesias2023X-shooterEvolution,
	author = {{Iglesias}, Daniela P. and {Pani{\'c}}, Olja and {van den Ancker}, Mario and {Petr-Gotzens}, Monika G. and {Siess}, Lionel and {Vioque}, Miguel and {Pascucci}, Ilaria and {Oudmaijer}, Ren{\'e} and {Miley}, James},
        title = "{X-shooter survey of young intermediate-mass stars - I. Stellar characterization and disc evolution}",
      journal = {\mnras},
         year = 2023,
        month = mar,
       volume = {519},
       number = {3},
        pages = {3958-3975},
          doi = {10.1093/mnras/stac3619},
archivePrefix = {arXiv},
       eprint = {2212.06791},
 primaryClass = {astro-ph.SR},
       adsurl = {https://ui.adsabs.harvard.edu/abs/2023MNRAS.519.3958I}
}

@article{Ishihara2017FaintIRSF,
	author = {{Ishihara}, Daisuke and {Takeuchi}, Nami and {Kobayashi}, Hiroshi and {Nagayama}, Takahiro and {Kaneda}, Hidehiro and {Inutsuka}, Shu-ichiro and {Fujiwara}, Hideaki and {Onaka}, Takashi},
        title = "{Faint warm debris disks around nearby bright stars explored by AKARI and IRSF}",
      journal = {\aap},
         year = 2017,
        month = may,
       volume = {601},
          eid = {A72},
        pages = {A72},
          doi = {10.1051/0004-6361/201526215},
archivePrefix = {arXiv},
       eprint = {1608.04480},
 primaryClass = {astro-ph.SR},
       adsurl = {https://ui.adsabs.harvard.edu/abs/2017A&A...601A..72I}
}

@article{Kaufer1999CommissioningLa-Silla.,
	author = {{Kaufer}, A. and {Stahl}, O. and {Tubbesing}, S. and {N{\o}rregaard}, P. and {Avila}, G. and {Francois}, P. and {Pasquini}, L. and {Pizzella}, A.},
        title = "{Commissioning FEROS, the new high-resolution spectrograph at La-Silla.}",
      journal = {The Messenger},
         year = 1999,
        month = mar,
       volume = {95},
        pages = {8-12},
       adsurl = {https://ui.adsabs.harvard.edu/abs/1999Msngr..95....8K}
}

@article{Kausch2015Molecfit:Correction,
	author = {{Kausch}, W. and {Noll}, S. and {Smette}, A. and {Kimeswenger}, S. and {Barden}, M. and {Szyszka}, C. and {Jones}, A.~M. and {Sana}, H. and {Horst}, H. and {Kerber}, F.},
        title = "{Molecfit: A general tool for telluric absorption correction. II. Quantitative evaluation on ESO-VLT/X-Shooterspectra}",
      journal = {\aap},
         year = 2015,
        month = apr,
       volume = {576},
          eid = {A78},
        pages = {A78},
          doi = {10.1051/0004-6361/201423909},
archivePrefix = {arXiv},
       eprint = {1501.07265},
 primaryClass = {astro-ph.IM},
       adsurl = {https://ui.adsabs.harvard.edu/abs/2015A&A...576A..78K}
}

@article{Kiefer2014TwoSystem,
	author = {{Kiefer}, F. and {Lecavelier des Etangs}, A. and {Boissier}, J. and {Vidal-Madjar}, A. and {Beust}, H. and {Lagrange}, A. -M. and {H{\'e}brard}, G. and {Ferlet}, R.},
        title = "{Two families of exocomets in the {\ensuremath{\beta}} Pictoris system}",
      journal = {\nat},
         year = 2014,
        month = oct,
       volume = {514},
       number = {7523},
        pages = {462-464},
          doi = {10.1038/nature13849},
       adsurl = {https://ui.adsabs.harvard.edu/abs/2014Natur.514..462K}
}

@article{Kiefer2014ExocometsHD172555,
	 author = {{Kiefer}, F. and {Lecavelier des Etangs}, A. and {Augereau}, J. -C. and {Vidal-Madjar}, A. and {Lagrange}, A. -M. and {Beust}, H.},
        title = "{Exocomets in the circumstellar gas disk of HD 172555}",
      journal = {\aap},
         year = 2014,
        month = jan,
       volume = {561},
          eid = {L10},
        pages = {L10},
          doi = {10.1051/0004-6361/201323128},
archivePrefix = {arXiv},
       eprint = {1401.1365},
 primaryClass = {astro-ph.EP},
       adsurl = {https://ui.adsabs.harvard.edu/abs/2014A&A...561L..10K}
}

@article{Kluska2022APlanets,
	author = {{Kluska}, J. and {Van Winckel}, H. and {Copp{\'e}e}, Q. and {Oomen}, G. -M. and {Dsilva}, K. and {Kamath}, D. and {Bujarrabal}, V. and {Min}, M.},
        title = "{A population of transition disks around evolved stars: Fingerprints of planets. Catalog of disks surrounding Galactic post-AGB binaries}",
      journal = {\aap},
         year = 2022,
        month = feb,
       volume = {658},
          eid = {A36},
        pages = {A36},
          doi = {10.1051/0004-6361/202141690},
archivePrefix = {arXiv},
       eprint = {2201.13155},
 primaryClass = {astro-ph.EP},
       adsurl = {https://ui.adsabs.harvard.edu/abs/2022A&A...658A..36K}
}

@article{Kospal2013ALMA21997,
	author = {{K{\'o}sp{\'a}l}, {\'A}. and {Mo{\'o}r}, A. and {Juh{\'a}sz}, A. and {{\'A}brah{\'a}m}, P. and {Apai}, D. and {Csengeri}, T. and {Grady}, C.~A. and {Henning}, Th. and {Hughes}, A.~M. and {Kiss}, Cs. and {Pascucci}, I. and {Schmalzl}, M.},
        title = "{ALMA Observations of the Molecular Gas in the Debris Disk of the 30 Myr Old Star HD 21997}",
      journal = {\apj},
         year = 2013,
        month = oct,
       volume = {776},
       number = {2},
          eid = {77},
        pages = {77},
          doi = {10.1088/0004-637X/776/2/77},
archivePrefix = {arXiv},
       eprint = {1310.5068},
 primaryClass = {astro-ph.SR},
       adsurl = {https://ui.adsabs.harvard.edu/abs/2013ApJ...776...77K}
}

@article{Koubsky1993ComingHerculis.,
	 author = {{Koubsky}, P. and {Horn}, J. and {Harmanec}, P. and {Hubert}, A.~M. and {Hubert}, H. and {Floquet}, M.},
        title = "{Coming shell phase of the Be star 4 Herculis.}",
      journal = {\aap},
         year = 1993,
        month = oct,
       volume = {277},
        pages = {521-523},
       adsurl = {https://ui.adsabs.harvard.edu/abs/1993A&A...277..521K}
}

@article{Kral2017PredictionsPlanetesimals,
	 author = {{Kral}, Quentin and {Matr{\`a}}, Luca and {Wyatt}, Mark C. and {Kennedy}, Grant M.},
        title = "{Predictions for the secondary CO, C and O gas content of debris discs from the destruction of volatile-rich planetesimals}",
      journal = {\mnras},
         year = 2017,
        month = jul,
       volume = {469},
       number = {1},
        pages = {521-550},
          doi = {10.1093/mnras/stx730},
archivePrefix = {arXiv},
       eprint = {1703.10693},
 primaryClass = {astro-ph.EP},
       adsurl = {https://ui.adsabs.harvard.edu/abs/2017MNRAS.469..521K}
}

@article{Kral2019ImagingDiscs,
	 author = {{Kral}, Quentin and {Marino}, Sebastian and {Wyatt}, Mark C. and {Kama}, Mihkel and {Matr{\`a}}, Luca},
        title = "{Imaging [CI] around HD 131835: reinterpreting young debris discs with protoplanetary disc levels of CO gas as shielded secondary discs}",
      journal = {\mnras},
         year = 2019,
        month = nov,
       volume = {489},
       number = {4},
        pages = {3670-3691},
          doi = {10.1093/mnras/sty2923},
archivePrefix = {arXiv},
       eprint = {1811.08439},
 primaryClass = {astro-ph.EP},
       adsurl = {https://ui.adsabs.harvard.edu/abs/2019MNRAS.489.3670K}
}

@article{Kunitomo2021PhotoevaporativeStars,
	author = {{Kunitomo}, Masanobu and {Ida}, Shigeru and {Takeuchi}, Taku and {Pani{\'c}}, Olja and {Miley}, James M. and {Suzuki}, Takeru K.},
        title = "{Photoevaporative Dispersal of Protoplanetary Disks around Evolving Intermediate-mass Stars}",
      journal = {\apj},
         year = 2021,
        month = mar,
       volume = {909},
       number = {2},
          eid = {109},
        pages = {109},
          doi = {10.3847/1538-4357/abdb2a},
archivePrefix = {arXiv},
       eprint = {2103.07673},
 primaryClass = {astro-ph.EP},
       adsurl = {https://ui.adsabs.harvard.edu/abs/2021ApJ...909..109K}
}

@article{Lagrange-Henri1990SearchStar.,
	author = {{Lagrange-Henri}, A.~M. and {Ferlet}, R. and {Vidal-Madjar}, A. and {Beust}, H. and {Gry}, C. and {Lallement}, R.},
        title = "{Search for beta Pictoris-like star.}",
      journal = {\aaps},
         year = 1990,
        month = nov,
       volume = {85},
        pages = {1089},
       adsurl = {https://ui.adsabs.harvard.edu/abs/1990A&AS...85.1089L}
}

@article{Lieman-Sifry2016DebrisALMA,
	author = {{Lieman-Sifry}, Jesse and {Hughes}, A. Meredith and {Carpenter}, John M. and {Gorti}, Uma and {Hales}, Antonio and {Flaherty}, Kevin M.},
        title = "{Debris Disks in the Scorpius-Centaurus OB Association Resolved by ALMA}",
      journal = {\apj},
         year = 2016,
        month = sep,
       volume = {828},
       number = {1},
          eid = {25},
        pages = {25},
          doi = {10.3847/0004-637X/828/1/25},
archivePrefix = {arXiv},
       eprint = {1606.07068},
 primaryClass = {astro-ph.EP},
       adsurl = {https://ui.adsabs.harvard.edu/abs/2016ApJ...828...25L}
}

@article{Lovell2021RapidDisc,
	author = {{Lovell}, J.~B. and {Kennedy}, G.~M. and {Marino}, S. and {Wyatt}, M.~C. and {Ansdell}, M. and {Kama}, M. and {Manara}, C.~F. and {Matr{\`a}}, L. and {Rosotti}, G. and {Tazzari}, M. and {Testi}, L. and {Williams}, J.~P.},
        title = "{Rapid CO gas dispersal from NO Lup's class III circumstellar disc}",
      journal = {\mnras},
         year = 2021,
        month = mar,
       volume = {502},
       number = {1},
        pages = {L66-L71},
          doi = {10.1093/mnrasl/slaa189},
archivePrefix = {arXiv},
       eprint = {2011.13229},
 primaryClass = {astro-ph.EP},
       adsurl = {https://ui.adsabs.harvard.edu/abs/2021MNRAS.502L..66L}
}

@article{Lovell2021ALMADispersal,
	 author = {{Lovell}, J.~B. and {Wyatt}, M.~C. and {Ansdell}, M. and {Kama}, M. and {Kennedy}, G.~M. and {Manara}, C.~F. and {Marino}, S. and {Matr{\`a}}, L. and {Rosotti}, G. and {Tazzari}, M. and {Testi}, L. and {Williams}, J.~P.},
        title = "{ALMA survey of Lupus class III stars: Early planetesimal belt formation and rapid disc dispersal}",
      journal = {\mnras},
         year = 2021,
        month = jan,
       volume = {500},
       number = {4},
        pages = {4878-4900},
          doi = {10.1093/mnras/staa3335},
archivePrefix = {arXiv},
       eprint = {2010.12657},
 primaryClass = {astro-ph.EP},
       adsurl = {https://ui.adsabs.harvard.edu/abs/2021MNRAS.500.4878L}
}

@article{MacGregor2017ADisk,
	author = {{MacGregor}, Meredith A. and {Matr{\`a}}, Luca and {Kalas}, Paul and {Wilner}, David J. and {Pan}, Margaret and {Kennedy}, Grant M. and {Wyatt}, Mark C. and {Duchene}, Gaspard and {Hughes}, A. Meredith and {Rieke}, George H. and {Clampin}, Mark and {Fitzgerald}, Michael P. and {Graham}, James R. and {Holland}, Wayne S. and {Pani{\'c}}, Olja and {Shannon}, Andrew and {Su}, Kate},
        title = "{A Complete ALMA Map of the Fomalhaut Debris Disk}",
      journal = {\apj},
         year = 2017,
        month = jun,
       volume = {842},
       number = {1},
          eid = {8},
        pages = {8},
          doi = {10.3847/1538-4357/aa71ae},
archivePrefix = {arXiv},
       eprint = {1705.05867},
 primaryClass = {astro-ph.EP},
       adsurl = {https://ui.adsabs.harvard.edu/abs/2017ApJ...842....8M}
}

@article{Malamut2014THEPc,
	 author = {{Malamut}, Craig and {Redfield}, Seth and {Linsky}, Jeffrey L. and {Wood}, Brian E. and {Ayres}, Thomas R.},
        title = "{The Structure of the Local Interstellar Medium. VI. New Mg II, Fe II, and Mn II Observations toward Stars within 100 pc}",
      journal = {\apj},
         year = 2014,
        month = may,
       volume = {787},
       number = {1},
          eid = {75},
        pages = {75},
          doi = {10.1088/0004-637X/787/1/75},
archivePrefix = {arXiv},
       eprint = {1403.8096},
 primaryClass = {astro-ph.SR},
       adsurl = {https://ui.adsabs.harvard.edu/abs/2014ApJ...787...75M}
}

@article{Marino2016ExocometaryRing,
	author = {{Marino}, S. and {Matr{\`a}}, L. and {Stark}, C. and {Wyatt}, M.~C. and {Casassus}, S. and {Kennedy}, G. and {Rodriguez}, D. and {Zuckerman}, B. and {Perez}, S. and {Dent}, W.~R.~F. and {Kuchner}, M. and {Hughes}, A.~M. and {Schneider}, G. and {Steele}, A. and {Roberge}, A. and {Donaldson}, J. and {Nesvold}, E.},
        title = "{Exocometary gas in the HD 181327 debris ring}",
      journal = {\mnras},
         year = 2016,
        month = aug,
       volume = {460},
       number = {3},
        pages = {2933-2944},
          doi = {10.1093/mnras/stw1216},
archivePrefix = {arXiv},
       eprint = {1605.05331},
 primaryClass = {astro-ph.EP},
       adsurl = {https://ui.adsabs.harvard.edu/abs/2016MNRAS.460.2933M}
}

@article{Marino2017ALMAPlanets,
	  author = {{Marino}, S. and {Wyatt}, M.~C. and {Pani{\'c}}, O. and {Matr{\`a}}, L. and {Kennedy}, G.~M. and {Bonsor}, A. and {Kral}, Q. and {Dent}, W.~R.~F. and {Duchene}, G. and {Wilner}, D. and {Lisse}, C.~M. and {Lestrade}, J. -F. and {Matthews}, B.},
        title = "{ALMA observations of the {\ensuremath{\eta}} Corvi debris disc: inward scattering of CO-rich exocomets by a chain of 3-30 M$_{{\ensuremath{\oplus}}}$ planets?}",
      journal = {\mnras},
         year = 2017,
        month = mar,
       volume = {465},
       number = {3},
        pages = {2595-2615},
          doi = {10.1093/mnras/stw2867},
archivePrefix = {arXiv},
       eprint = {1611.01168},
 primaryClass = {astro-ph.EP},
       adsurl = {https://ui.adsabs.harvard.edu/abs/2017MNRAS.465.2595M}
}

@article{Marino2022VerticalLifetime,
	author = {{Marino}, S. and {Cataldi}, G. and {Jankovic}, M.~R. and {Matr{\`a}}, L. and {Wyatt}, M.~C.},
        title = "{Vertical evolution of exocometary gas - I. How vertical diffusion shortens the CO lifetime}",
      journal = {\mnras},
         year = 2022,
        month = sep,
       volume = {515},
       number = {1},
        pages = {507-524},
          doi = {10.1093/mnras/stac1756},
archivePrefix = {arXiv},
       eprint = {2206.11071},
 primaryClass = {astro-ph.EP},
       adsurl = {https://ui.adsabs.harvard.edu/abs/2022MNRAS.515..507M}
}

@article{Matra2015COALMA,
	author = {{Matr{\`a}}, L. and {Pani{\'c}}, O. and {Wyatt}, M.~C. and {Dent}, W.~R.~F.},
        title = "{CO mass upper limits in the Fomalhaut ring - the importance of NLTE excitation in debris discs and future prospects with ALMA}",
      journal = {\mnras},
         year = 2015,
        month = mar,
       volume = {447},
       number = {4},
        pages = {3936-3947},
          doi = {10.1093/mnras/stu2619},
archivePrefix = {arXiv},
       eprint = {1412.2757},
 primaryClass = {astro-ph.EP},
       adsurl = {https://ui.adsabs.harvard.edu/abs/2015MNRAS.447.3936M}
}

@article{Matra2017DetectionComets,
	author = {{Matr{\`a}}, L. and {MacGregor}, M.~A. and {Kalas}, P. and {Wyatt}, M.~C. and {Kennedy}, G.~M. and {Wilner}, D.~J. and {Duchene}, G. and {Hughes}, A.~M. and {Pan}, M. and {Shannon}, A. and {Clampin}, M. and {Fitzgerald}, M.~P. and {Graham}, J.~R. and {Holland}, W.~S. and {Pani{\'c}}, O. and {Su}, K.~Y.~L.},
        title = "{Detection of Exocometary CO within the 440 Myr Old Fomalhaut Belt: A Similar CO+CO$_{2}$ Ice Abundance in Exocomets and Solar System Comets}",
      journal = {\apj},
         year = 2017,
        month = jun,
       volume = {842},
       number = {1},
          eid = {9},
        pages = {9},
          doi = {10.3847/1538-4357/aa71b4},
archivePrefix = {arXiv},
       eprint = {1705.05868},
 primaryClass = {astro-ph.EP},
       adsurl = {https://ui.adsabs.harvard.edu/abs/2017ApJ...842....9M}
}

@article{Matra2017ExocometaryObservations,
	author = {{Matr{\`a}}, L. and {Dent}, W.~R.~F. and {Wyatt}, M.~C. and {Kral}, Q. and {Wilner}, D.~J. and {Pani{\'c}}, O. and {Hughes}, A.~M. and {de Gregorio-Monsalvo}, I. and {Hales}, A. and {Augereau}, J. -C. and {Greaves}, J. and {Roberge}, A.},
        title = "{Exocometary gas structure, origin and physical properties around {\ensuremath{\beta}} Pictoris through ALMA CO multitransition observations}",
      journal = {\mnras},
         year = 2017,
        month = jan,
       volume = {464},
       number = {2},
        pages = {1415-1433},
          doi = {10.1093/mnras/stw2415},
archivePrefix = {arXiv},
       eprint = {1609.06718},
 primaryClass = {astro-ph.EP},
       adsurl = {https://ui.adsabs.harvard.edu/abs/2017MNRAS.464.1415M}
}

@article{Matra2019On7,
	 author = {{Matr{\`a}}, L. and {{\"O}berg}, K.~I. and {Wilner}, D.~J. and {Olofsson}, J. and {Bayo}, A.},
        title = "{On the Ubiquity and Stellar Luminosity Dependence of Exocometary CO Gas: Detection around M Dwarf TWA 7}",
      journal = {\aj},
         year = 2019,
        month = mar,
       volume = {157},
       number = {3},
          eid = {117},
        pages = {117},
          doi = {10.3847/1538-3881/aaff5b},
archivePrefix = {arXiv},
       eprint = {1901.05004},
 primaryClass = {astro-ph.EP},
       adsurl = {https://ui.adsabs.harvard.edu/abs/2019AJ....157..117M}
}

@article{Mayor2003SettingHARPS,
	 author = {{Mayor}, M. and {Pepe}, F. and {Queloz}, D. and {Bouchy}, F. and {Rupprecht}, G. and {Lo Curto}, G. and {Avila}, G. and {Benz}, W. and {Bertaux}, J. -L. and {Bonfils}, X. and {Dall}, Th. and {Dekker}, H. and {Delabre}, B. and {Eckert}, W. and {Fleury}, M. and {Gilliotte}, A. and {Gojak}, D. and {Guzman}, J.~C. and {Kohler}, D. and {Lizon}, J. -L. and {Longinotti}, A. and {Lovis}, C. and {Megevand}, D. and {Pasquini}, L. and {Reyes}, J. and {Sivan}, J. -P. and {Sosnowska}, D. and {Soto}, R. and {Udry}, S. and {van Kesteren}, A. and {Weber}, L. and {Weilenmann}, U.},
        title = "{Setting New Standards with HARPS}",
      journal = {The Messenger},
         year = 2003,
        month = dec,
       volume = {114},
        pages = {20-24},
       adsurl = {https://ui.adsabs.harvard.edu/abs/2003Msngr.114...20M}
}

@inproceedings{McMullin2007CASAApplications,
	author = {{McMullin}, J.~P. and {Waters}, B. and {Schiebel}, D. and {Young}, W. and {Golap}, K.},
        title = "{CASA Architecture and Applications}",
    booktitle = {Astronomical Data Analysis Software and Systems XVI},
         year = 2007,
       editor = {{Shaw}, R.~A. and {Hill}, F. and {Bell}, D.~J.},
       series = {Astronomical Society of the Pacific Conference Series},
       volume = {376},
        month = oct,
        pages = {127},
       adsurl = {https://ui.adsabs.harvard.edu/abs/2007ASPC..376..127M}
}

@article{Miley2018Unlocking141569,
	author = {{Miley}, J.~M. and {Pani{\'c}}, O. and {Wyatt}, M. and {Kennedy}, G.~M.},
        title = "{Unlocking the secrets of the midplane gas and dust distribution in the young hybrid disc HD 141569}",
      journal = {\aap},
         year = 2018,
        month = jul,
       volume = {615},
          eid = {L10},
        pages = {L10},
          doi = {10.1051/0004-6361/201833381},
archivePrefix = {arXiv},
       eprint = {1805.02476},
 primaryClass = {astro-ph.EP},
       adsurl = {https://ui.adsabs.harvard.edu/abs/2018A&A...615L..10M}
}

@article{Miley2019Asymmetric100546,
	author = {{Miley}, J.~M. and {Pani{\'c}}, O. and {Haworth}, T.~J. and {Pascucci}, I. and {Wyatt}, M. and {Clarke}, C. and {Richards}, A.~M.~S. and {Ratzka}, T.},
        title = "{Asymmetric mid-plane gas in ALMA images of HD 100546}",
      journal = {\mnras},
         year = 2019,
        month = may,
       volume = {485},
       number = {1},
        pages = {739-752},
          doi = {10.1093/mnras/stz426},
archivePrefix = {arXiv},
       eprint = {1902.07506},
 primaryClass = {astro-ph.SR},
       adsurl = {https://ui.adsabs.harvard.edu/abs/2019MNRAS.485..739M}
}

@article{Montgomery2012DetectionStars,
	 author = {{Montgomery}, Sharon L. and {Welsh}, Barry Y.},
        title = "{Detection of Variable Gaseous Absorption Features in the Debris Disks Around Young A-type Stars}",
      journal = {\pasp},
         year = 2012,
        month = oct,
       volume = {124},
       number = {920},
        pages = {1042},
          doi = {10.1086/668293},
       adsurl = {https://ui.adsabs.harvard.edu/abs/2012PASP..124.1042M}
}

@article{Moor2013ALMA21997,
	author = {{Mo{\'o}r}, A. and {Juh{\'a}sz}, A. and {K{\'o}sp{\'a}l}, {\'A}. and {{\'A}brah{\'a}m}, P. and {Apai}, D. and {Csengeri}, T. and {Grady}, C. and {Henning}, Th. and {Hughes}, A.~M. and {Kiss}, Cs. and {Pascucci}, I. and {Schmalzl}, M. and {Gab{\'a}nyi}, K.},
        title = "{ALMA Continuum Observations of a 30 Myr Old Gaseous Debris Disk around HD 21997}",
      journal = {\apjl},
         year = 2013,
        month = nov,
       volume = {777},
       number = {2},
          eid = {L25},
        pages = {L25},
          doi = {10.1088/2041-8205/777/2/L25},
archivePrefix = {arXiv},
       eprint = {1310.5069},
 primaryClass = {astro-ph.SR},
       adsurl = {https://ui.adsabs.harvard.edu/abs/2013ApJ...777L..25M}
}

@article{Moor2017MolecularStars,
	author = {{Mo{\'o}r}, Attila and {Cur{\'e}}, Michel and {K{\'o}sp{\'a}l}, {\'A}gnes and {{\'A}brah{\'a}m}, P{\'e}ter and {Csengeri}, Timea and {Eiroa}, Carlos and {Gunawan}, Diah and {Henning}, Thomas and {Hughes}, A. Meredith and {Juh{\'a}sz}, Attila and {Pawellek}, Nicole and {Wyatt}, Mark},
        title = "{Molecular Gas in Debris Disks around Young A-type Stars}",
      journal = {\apj},
         year = 2017,
        month = nov,
       volume = {849},
       number = {2},
          eid = {123},
        pages = {123},
          doi = {10.3847/1538-4357/aa8e4e},
archivePrefix = {arXiv},
       eprint = {1709.08414},
 primaryClass = {astro-ph.SR},
       adsurl = {https://ui.adsabs.harvard.edu/abs/2017ApJ...849..123M}
}

@article{Moor2019New32297,
	author = {{Mo{\'o}r}, Attila and {Kral}, Quentin and {{\'A}brah{\'a}m}, P{\'e}ter and {K{\'o}sp{\'a}l}, {\'A}gnes and {Dutrey}, Anne and {Di Folco}, Emmanuel and {Hughes}, A. Meredith and {Juh{\'a}sz}, Attila and {Pascucci}, Ilaria and {Pawellek}, Nicole},
        title = "{New Millimeter CO Observations of the Gas-rich Debris Disks 49 Cet and HD 32297}",
      journal = {\apj},
         year = 2019,
        month = oct,
       volume = {884},
       number = {2},
          eid = {108},
        pages = {108},
          doi = {10.3847/1538-4357/ab4272},
archivePrefix = {arXiv},
       eprint = {1908.09685},
 primaryClass = {astro-ph.EP},
       adsurl = {https://ui.adsabs.harvard.edu/abs/2019ApJ...884..108M}
}

@article{Mulders2013WhyHD100546,
	 author = {{Mulders}, G.~D. and {Min}, M. and {Dominik}, C. and {Debes}, J.~H. and {Schneider}, G.},
        title = "{Why circumstellar disks are so faint in scattered light: the case of HD 100546}",
      journal = {\aap},
         year = 2013,
        month = jan,
       volume = {549},
          eid = {A112},
        pages = {A112},
          doi = {10.1051/0004-6361/201219522},
archivePrefix = {arXiv},
       eprint = {1210.4132},
 primaryClass = {astro-ph.SR},
       adsurl = {https://ui.adsabs.harvard.edu/abs/2013A&A...549A.112M}
}

@article{Nakatani2021PhotoevaporationRemnants,
	author = {{Nakatani}, Riouhei and {Kobayashi}, Hiroshi and {Kuiper}, Rolf and {Nomura}, Hideko and {Aikawa}, Yuri},
        title = "{Photoevaporation of Grain-depleted Protoplanetary Disks around Intermediate-mass Stars: Investigating the Possibility of Gas-rich Debris Disks as Protoplanetary Remnants}",
      journal = {\apj},
         year = 2021,
        month = jul,
       volume = {915},
       number = {2},
          eid = {90},
        pages = {90},
          doi = {10.3847/1538-4357/ac0137},
archivePrefix = {arXiv},
       eprint = {2009.06438},
 primaryClass = {astro-ph.SR},
       adsurl = {https://ui.adsabs.harvard.edu/abs/2021ApJ...915...90N}
}

@article{Panic2017EffectsLocation,
	author = {{Pani{\'c}}, O. and {Min}, M.},
        title = "{Effects of disc mid-plane evolution on CO snowline location}",
      journal = {\mnras},
         year = 2017,
        month = may,
       volume = {467},
       number = {1},
        pages = {1175-1185},
          doi = {10.1093/mnras/stx114},
archivePrefix = {arXiv},
       eprint = {1703.09708},
 primaryClass = {astro-ph.EP},
       adsurl = {https://ui.adsabs.harvard.edu/abs/2017MNRAS.467.1175P}
}

@article{Panic2013FirstDiscs,
	author = {{Pani{\'c}}, O. and {Holland}, W.~S. and {Wyatt}, M.~C. and {Kennedy}, G.~M. and {Matthews}, B.~C. and {Lestrade}, J.~F. and {Sibthorpe}, B. and {Greaves}, J.~S. and {Marshall}, J.~P. and {Phillips}, N.~M. and {Tottle}, J.},
        title = "{First results of the SONS survey: submillimetre detections of debris discs}",
      journal = {\mnras},
         year = 2013,
        month = oct,
       volume = {435},
       number = {2},
        pages = {1037-1046},
          doi = {10.1093/mnras/stt1293},
       adsurl = {https://ui.adsabs.harvard.edu/abs/2013MNRAS.435.1037P}
}

@ARTICLE{Pearce2021Fomalhaut,
       author = {{Pearce}, Tim D. and {Beust}, Herv{\'e} and {Faramaz}, Virginie and {Booth}, Mark and {Krivov}, Alexander V. and {L{\"o}hne}, Torsten and {Poblete}, Pedro P.},
        title = "{Fomalhaut b could be massive and sculpting the narrow, eccentric debris disc, if in mean-motion resonance with it}",
      journal = {\mnras},
         year = 2021,
        month = jun,
       volume = {503},
       number = {4},
        pages = {4767-4786},
          doi = {10.1093/mnras/stab760},
archivePrefix = {arXiv},
       eprint = {2103.04977},
       adsurl = {https://ui.adsabs.harvard.edu/abs/2021MNRAS.503.4767P}
}

@article{Pearce2022PlanetDiscs,
	 author = {{Pearce}, Tim D. and {Launhardt}, Ralf and {Ostermann}, Robert and {Kennedy}, Grant M. and {Gennaro}, Mario and {Booth}, Mark and {Krivov}, Alexander V. and {Cugno}, Gabriele and {Henning}, Thomas K. and {Quirrenbach}, Andreas and {Barcucci}, Arianna Musso and {Matthews}, Elisabeth C. and {Ruh}, Henrik L. and {Stone}, Jordan M.},
        title = "{Planet populations inferred from debris discs. Insights from 178 debris systems in the ISPY, LEECH, and LIStEN planet-hunting surveys}",
      journal = {\aap},
         year = 2022,
        month = mar,
       volume = {659},
          eid = {A135},
        pages = {A135},
          doi = {10.1051/0004-6361/202142720},
archivePrefix = {arXiv},
       eprint = {2201.08369},
 primaryClass = {astro-ph.EP},
       adsurl = {https://ui.adsabs.harvard.edu/abs/2022A&A...659A.135P}
}

@article{Pericaud2017TheDisks,
	author = {{P{\'e}ricaud}, J. and {Di Folco}, E. and {Dutrey}, A. and {Guilloteau}, S. and {Pi{\'e}tu}, V.},
        title = "{The hybrid disks: a search and study to better understand evolution of disks}",
      journal = {\aap},
         year = 2017,
        month = apr,
       volume = {600},
          eid = {A62},
        pages = {A62},
          doi = {10.1051/0004-6361/201629371},
archivePrefix = {arXiv},
       eprint = {1612.06582},
 primaryClass = {astro-ph.SR},
       adsurl = {https://ui.adsabs.harvard.edu/abs/2017A&A...600A..62P}
}

@article{Prusti2016TheMission,
	author = {{Gaia Collaboration} and {Prusti}, T. and {de Bruijne}, J.~H.~J. and {Brown}, A.~G.~A. and {Vallenari}, A. and {Babusiaux}, C. and {Bailer-Jones}, C.~A.~L. and {Bastian}, U. and {Biermann}, M. and {Evans}, D.~W. and {Eyer}, L. and {Jansen}, F. and {Jordi}, C. and {Klioner}, S.~A. and {Lammers}, U. and {Lindegren}, L. and {Luri}, X. and {Mignard}, F. and {Milligan}, D.~J. and {Panem}, C. and {Poinsignon}, V. and {Pourbaix}, D. and {Randich}, S. and {Sarri}, G. and {Sartoretti}, P. and {Siddiqui}, H.~I. and {Soubiran}, C. and {Valette}, V. and {van Leeuwen}, F. and {Walton}, N.~A. and {Aerts}, C. and {Arenou}, F. and {Cropper}, M. and {Drimmel}, R. and {H{\o}g}, E. and {Katz}, D. and {Lattanzi}, M.~G. and {O'Mullane}, W. and {Grebel}, E.~K. and {Holland}, A.~D. and {Huc}, C. and {Passot}, X. and {Bramante}, L. and {Cacciari}, C. and {Casta{\~n}eda}, J. and {Chaoul}, L. and {Cheek}, N. and {De Angeli}, F. and {Fabricius}, C. and {Guerra}, R. and {Hern{\'a}ndez}, J. and {Jean-Antoine-Piccolo}, A. and {Masana}, E. and {Messineo}, R. and {Mowlavi}, N. and {Nienartowicz}, K. and {Ord{\'o}{\~n}ez-Blanco}, D. and {Panuzzo}, P. and {Portell}, J. and {Richards}, P.~J. and {Riello}, M. and {Seabroke}, G.~M. and {Tanga}, P. and {Th{\'e}venin}, F. and {Torra}, J. and {Els}, S.~G. and {Gracia-Abril}, G. and {Comoretto}, G. and {Garcia-Reinaldos}, M. and {Lock}, T. and {Mercier}, E. and {Altmann}, M. and {Andrae}, R. and {Astraatmadja}, T.~L. and {Bellas-Velidis}, I. and {Benson}, K. and {Berthier}, J. and {Blomme}, R. and {Busso}, G. and {Carry}, B. and {Cellino}, A. and {Clementini}, G. and {Cowell}, S. and {Creevey}, O. and {Cuypers}, J. and {Davidson}, M. and {De Ridder}, J. and {de Torres}, A. and {Delchambre}, L. and {Dell'Oro}, A. and {Ducourant}, C. and {Fr{\'e}mat}, Y. and {Garc{\'\i}a-Torres}, M. and {Gosset}, E. and {Halbwachs}, J. -L. and {Hambly}, N.~C. and {Harrison}, D.~L. and {Hauser}, M. and {Hestroffer}, D. and {Hodgkin}, S.~T. and {Huckle}, H.~E. and {Hutton}, A. and {Jasniewicz}, G. and {Jordan}, S. and {Kontizas}, M. and {Korn}, A.~J. and {Lanzafame}, A.~C. and {Manteiga}, M. and {Moitinho}, A. and {Muinonen}, K. and {Osinde}, J. and {Pancino}, E. and {Pauwels}, T. and {Petit}, J. -M. and {Recio-Blanco}, A. and {Robin}, A.~C. and {Sarro}, L.~M. and {Siopis}, C. and {Smith}, M. and {Smith}, K.~W. and {Sozzetti}, A. and {Thuillot}, W. and {van Reeven}, W. and {Viala}, Y. and {Abbas}, U. and {Abreu Aramburu}, A. and {Accart}, S. and {Aguado}, J.~J. and {Allan}, P.~M. and {Allasia}, W. and {Altavilla}, G. and {{\'A}lvarez}, M.~A. and {Alves}, J. and {Anderson}, R.~I. and {Andrei}, A.~H. and {Anglada Varela}, E. and {Antiche}, E. and {Antoja}, T. and {Ant{\'o}n}, S. and {Arcay}, B. and {Atzei}, A. and {Ayache}, L. and {Bach}, N. and {Baker}, S.~G. and {Balaguer-N{\'u}{\~n}ez}, L. and {Barache}, C. and {Barata}, C. and {Barbier}, A. and {Barblan}, F. and {Baroni}, M. and {Barrado y Navascu{\'e}s}, D. and {Barros}, M. and {Barstow}, M.~A. and {Becciani}, U. and {Bellazzini}, M. and {Bellei}, G. and {Bello Garc{\'\i}a}, A. and {Belokurov}, V. and {Bendjoya}, P. and {Berihuete}, A. and {Bianchi}, L. and {Bienaym{\'e}}, O. and {Billebaud}, F. and {Blagorodnova}, N. and {Blanco-Cuaresma}, S. and {Boch}, T. and {Bombrun}, A. and {Borrachero}, R. and {Bouquillon}, S. and {Bourda}, G. and {Bouy}, H. and {Bragaglia}, A. and {Breddels}, M.~A. and {Brouillet}, N. and {Br{\"u}semeister}, T. and {Bucciarelli}, B. and {Budnik}, F. and {Burgess}, P. and {Burgon}, R. and {Burlacu}, A. and {Busonero}, D. and {Buzzi}, R. and {Caffau}, E. and {Cambras}, J. and {Campbell}, H. and {Cancelliere}, R. and {Cantat-Gaudin}, T. and {Carlucci}, T. and {Carrasco}, J.~M. and {Castellani}, M. and {Charlot}, P. and {Charnas}, J. and {Charvet}, P. and {Chassat}, F. and {Chiavassa}, A. and {Clotet}, M. and {Cocozza}, G. and {Collins}, R.~S. and {Collins}, P. and {Costigan}, G.},
        title = "{The Gaia mission}",
      journal = {\aap},
         year = 2016,
        month = nov,
       volume = {595},
          eid = {A1},
        pages = {A1},
          doi = {10.1051/0004-6361/201629272},
archivePrefix = {arXiv},
       eprint = {1609.04153},
 primaryClass = {astro-ph.IM},
       adsurl = {https://ui.adsabs.harvard.edu/abs/2016A&A...595A...1G}
}

@article{Rebollido2018TheDiscs,
	author = {{Rebollido}, I. and {Eiroa}, C. and {Montesinos}, B. and {Maldonado}, J. and {Villaver}, E. and {Absil}, O. and {Bayo}, A. and {Canovas}, H. and {Carmona}, A. and {Chen}, Ch. and {Ertel}, S. and {Garufi}, A. and {Henning}, Th. and {Iglesias}, D.~P. and {Launhardt}, R. and {Liseau}, R. and {Meeus}, G. and {Mo{\'o}r}, A. and {Mora}, A. and {Olofsson}, J. and {Rauw}, G. and {Riviere-Marichalar}, P.},
        title = "{The co-existence of hot and cold gas in debris discs}",
      journal = {\aap},
         year = 2018,
        month = jun,
       volume = {614},
          eid = {A3},
        pages = {A3},
          doi = {10.1051/0004-6361/201732329},
archivePrefix = {arXiv},
       eprint = {1801.07951},
 primaryClass = {astro-ph.SR},
       adsurl = {https://ui.adsabs.harvard.edu/abs/2018A&A...614A...3R}
}

@article{Rebollido2020Exocomets:Survey,
	author = {{Rebollido}, I. and {Eiroa}, C. and {Montesinos}, B. and {Maldonado}, J. and {Villaver}, E. and {Absil}, O. and {Bayo}, A. and {Canovas}, H. and {Carmona}, A. and {Chen}, Ch. and {Ertel}, S. and {Henning}, Th. and {Iglesias}, D.~P. and {Launhardt}, R. and {Liseau}, R. and {Meeus}, G. and {Mo{\'o}r}, A. and {Mora}, A. and {Olofsson}, J. and {Rauw}, G. and {Riviere-Marichalar}, P.},
        title = "{Exocomets: A spectroscopic survey}",
      journal = {\aap},
         year = 2020,
        month = jul,
       volume = {639},
          eid = {A11},
        pages = {A11},
          doi = {10.1051/0004-6361/201936071},
archivePrefix = {arXiv},
       eprint = {2003.11084},
 primaryClass = {astro-ph.SR},
       adsurl = {https://ui.adsabs.harvard.edu/abs/2020A&A...639A..11R}
}

@article{Rebollido2022The36546,
	author = {{Rebollido}, Isabel and {Ribas}, {\'A}lvaro and {de Gregorio-Monsalvo}, Itziar and {Villaver}, Eva and {Montesinos}, Benjam{\'\i}n and {Chen}, Christine and {Canovas}, H{\'e}ctor and {Henning}, Thomas and {Mo{\'o}r}, Attila and {Perrin}, Marshall and {Rivi{\`e}re-Marichalar}, Pablo and {Eiroa}, Carlos},
        title = "{The search for gas in debris discs: ALMA detection of CO gas in HD 36546}",
      journal = {\mnras},
         year = 2022,
        month = jan,
       volume = {509},
       number = {1},
        pages = {693-700},
          doi = {10.1093/mnras/stab2906},
archivePrefix = {arXiv},
       eprint = {2110.02308},
 primaryClass = {astro-ph.EP},
       adsurl = {https://ui.adsabs.harvard.edu/abs/2022MNRAS.509..693R}
}

@article{Redfield2007Gas32297,
	author = {{Redfield}, Seth},
        title = "{Gas Absorption Detected from the Edge-on Debris Disk Surrounding HD 32297}",
      journal = {\apjl},
         year = 2007,
        month = feb,
       volume = {656},
       number = {2},
        pages = {L97-L100},
          doi = {10.1086/512237},
archivePrefix = {arXiv},
       eprint = {astro-ph/0701116},
 primaryClass = {astro-ph},
       adsurl = {https://ui.adsabs.harvard.edu/abs/2007ApJ...656L..97R}
}

@article{Riviere-Marichalar2014GasObservatory,
	author = {{Riviere-Marichalar}, P. and {Barrado}, D. and {Montesinos}, B. and {Duch{\^e}ne}, G. and {Bouy}, H. and {Pinte}, C. and {Menard}, F. and {Donaldson}, J. and {Eiroa}, C. and {Krivov}, A.~V. and {Kamp}, I. and {Mendigut{\'\i}a}, I. and {Dent}, W.~R.~F. and {Lillo-Box}, J.},
        title = "{Gas and dust in the beta Pictoris moving group as seen by the Herschel Space Observatory}",
      journal = {\aap},
         year = 2014,
        month = may,
       volume = {565},
          eid = {A68},
        pages = {A68},
          doi = {10.1051/0004-6361/201322901},
archivePrefix = {arXiv},
       eprint = {1404.1815},
 primaryClass = {astro-ph.SR},
       adsurl = {https://ui.adsabs.harvard.edu/abs/2014A&A...565A..68R}
}

@article{Roberge2014VOLATILE-RICHDISK,
	author = {{Roberge}, Aki and {Welsh}, Barry Y. and {Kamp}, Inga and {Weinberger}, Alycia J. and {Grady}, Carol A.},
        title = "{Volatile-rich Circumstellar Gas in the Unusual 49 Ceti Debris Disk}",
      journal = {\apjl},
         year = 2014,
        month = nov,
       volume = {796},
       number = {1},
          eid = {L11},
        pages = {L11},
          doi = {10.1088/2041-8205/796/1/L11},
archivePrefix = {arXiv},
       eprint = {1410.6542},
 primaryClass = {astro-ph.EP},
       adsurl = {https://ui.adsabs.harvard.edu/abs/2014ApJ...796L..11R}
}

@article{Sandell2011ASTARS,
	author = {{Sandell}, G{\"o}ran and {Weintraub}, David A. and {Hamidouche}, Murad},
        title = "{A Submillimeter Mapping Survey of Herbig AeBe Stars}",
      journal = {\apj},
         year = 2011,
        month = jan,
       volume = {727},
       number = {1},
          eid = {26},
        pages = {26},
          doi = {10.1088/0004-637X/727/1/26},
archivePrefix = {arXiv},
       eprint = {1011.3747},
 primaryClass = {astro-ph.SR},
       adsurl = {https://ui.adsabs.harvard.edu/abs/2011ApJ...727...26S}
}

@article{Sheret2004SubmillimetreStars,
	author = {{Sheret}, I. and {Dent}, W.~R.~F. and {Wyatt}, M.~C.},
        title = "{Submillimetre observations and modelling of Vega-type stars}",
      journal = {\mnras},
         year = 2004,
        month = mar,
       volume = {348},
       number = {4},
        pages = {1282-1294},
          doi = {10.1111/j.1365-2966.2004.07448.x},
archivePrefix = {arXiv},
       eprint = {astro-ph/0311593},
 primaryClass = {astro-ph},
       adsurl = {https://ui.adsabs.harvard.edu/abs/2004MNRAS.348.1282S}
}

@article{Smette2015Molecfit:Correction,
	author = {{Smette}, A. and {Sana}, H. and {Noll}, S. and {Horst}, H. and {Kausch}, W. and {Kimeswenger}, S. and {Barden}, M. and {Szyszka}, C. and {Jones}, A.~M. and {Gallenne}, A. and {Vinther}, J. and {Ballester}, P. and {Taylor}, J.},
        title = "{Molecfit: A general tool for telluric absorption correction. I. Method and application to ESO instruments}",
      journal = {\aap},
         year = 2015,
        month = apr,
       volume = {576},
          eid = {A77},
        pages = {A77},
          doi = {10.1051/0004-6361/201423932},
archivePrefix = {arXiv},
       eprint = {1501.07239},
 primaryClass = {astro-ph.IM},
       adsurl = {https://ui.adsabs.harvard.edu/abs/2015A&A...576A..77S}
}

@article{Smirnov-Pinchukov2022LackGas,
	author = {{Smirnov-Pinchukov}, Grigorii V. and {Mo{\'o}r}, Attila and {Semenov}, Dmitry A. and {{\'A}brah{\'a}m}, P{\'e}ter and {Henning}, Thomas and {K{\'o}sp{\'a}l}, {\'A}gnes and {Hughes}, A. Meredith and {di Folco}, Emmanuel},
        title = "{Lack of other molecules in CO-rich debris discs: is it primordial or secondary gas?}",
      journal = {\mnras},
         year = 2022,
        month = feb,
       volume = {510},
       number = {1},
        pages = {1148-1162},
          doi = {10.1093/mnras/stab3146},
archivePrefix = {arXiv},
       eprint = {2111.07655},
 primaryClass = {astro-ph.EP},
       adsurl = {https://ui.adsabs.harvard.edu/abs/2022MNRAS.510.1148S}
}

@article{Southworth2022ScutiHD23642,
	author = {{Southworth}, John and {Murphy}, S.~J. and {Pavlovski}, K.},
        title = "{{\ensuremath{\delta}} Scuti pulsations in the bright Pleiades eclipsing binary HD 23642}",
      journal = {\mnras},
         year = 2023,
        month = mar,
       volume = {520},
       number = {1},
        pages = {L53-L57},
          doi = {10.1093/mnrasl/slad004},
archivePrefix = {arXiv},
       eprint = {2301.04912},
 primaryClass = {astro-ph.SR},
       adsurl = {https://ui.adsabs.harvard.edu/abs/2023MNRAS.520L..53S}
}

@article{Stapper2022TheALMA,
	author = {{Stapper}, L.~M. and {Hogerheijde}, M.~R. and {van Dishoeck}, E.~F. and {Mentel}, R.},
        title = "{The mass and size of Herbig disks as seen by ALMA}",
      journal = {\aap},
         year = 2022,
        month = feb,
       volume = {658},
          eid = {A112},
        pages = {A112},
          doi = {10.1051/0004-6361/202142164},
archivePrefix = {arXiv},
       eprint = {2112.03297},
 primaryClass = {astro-ph.EP},
       adsurl = {https://ui.adsabs.harvard.edu/abs/2022A&A...658A.112S}
}

@article{Su2017ALMASystem,
	author = {{Su}, Kate Y.~L. and {MacGregor}, Meredith A. and {Booth}, Mark and {Wilner}, David J. and {Flaherty}, Kevin and {Hughes}, A. Meredith and {Phillips}, Neil M. and {Malhotra}, Renu and {Hales}, Antonio S. and {Morrison}, Sarah and {Ertel}, Steve and {Matthews}, Brenda C. and {Dent}, William R.~F. and {Casassus}, Simon},
        title = "{ALMA 1.3 mm Map of the HD 95086 System}",
      journal = {\aj},
         year = 2017,
        month = dec,
       volume = {154},
       number = {6},
          eid = {225},
        pages = {225},
          doi = {10.3847/1538-3881/aa906b},
archivePrefix = {arXiv},
       eprint = {1709.10129},
 primaryClass = {astro-ph.EP},
       adsurl = {https://ui.adsabs.harvard.edu/abs/2017AJ....154..225S}
}

@article{Vallenari2023iGaia/i3,
       author = {{Gaia Collaboration} and {Vallenari}, A. and {Brown}, A.~G.~A. and {Prusti}, T. and {de Bruijne}, J.~H.~J. and {Arenou}, F. and {Babusiaux}, C. and {Biermann}, M. and {Creevey}, O.~L. and {Ducourant}, C. and {Evans}, D.~W. and {Eyer}, L. and {Guerra}, R. and {Hutton}, A. and {Jordi}, C. and {Klioner}, S.~A. and {Lammers}, U.~L. and {Lindegren}, L. and {Luri}, X. and {Mignard}, F. and {Panem}, C. and {Pourbaix}, D. and {Randich}, S. and {Sartoretti}, P. and {Soubiran}, C. and {Tanga}, P. and {Walton}, N.~A. and {Bailer-Jones}, C.~A.~L. and {Bastian}, U. and {Drimmel}, R. and {Jansen}, F. and {Katz}, D. and {Lattanzi}, M.~G. and {van Leeuwen}, F. and {Bakker}, J. and {Cacciari}, C. and {Casta{\~n}eda}, J. and {De Angeli}, F. and {Fabricius}, C. and {Fouesneau}, M. and {Fr{\'e}mat}, Y. and {Galluccio}, L. and {Guerrier}, A. and {Heiter}, U. and {Masana}, E. and {Messineo}, R. and {Mowlavi}, N. and {Nicolas}, C. and {Nienartowicz}, K. and {Pailler}, F. and {Panuzzo}, P. and {Riclet}, F. and {Roux}, W. and {Seabroke}, G.~M. and {Sordo}, R. and {Th{\'e}venin}, F. and {Gracia-Abril}, G. and {Portell}, J. and {Teyssier}, D. and {Altmann}, M. and {Andrae}, R. and {Audard}, M. and {Bellas-Velidis}, I. and {Benson}, K. and {Berthier}, J. and {Blomme}, R. and {Burgess}, P.~W. and {Busonero}, D. and {Busso}, G. and {C{\'a}novas}, H. and {Carry}, B. and {Cellino}, A. and {Cheek}, N. and {Clementini}, G. and {Damerdji}, Y. and {Davidson}, M. and {de Teodoro}, P. and {Nu{\~n}ez Campos}, M. and {Delchambre}, L. and {Dell'Oro}, A. and {Esquej}, P. and {Fern{\'a}ndez-Hern{\'a}ndez}, J. and {Fraile}, E. and {Garabato}, D. and {Garc{\'\i}a-Lario}, P. and {Gosset}, E. and {Haigron}, R. and {Halbwachs}, J. -L. and {Hambly}, N.~C. and {Harrison}, D.~L. and {Hern{\'a}ndez}, J. and {Hestroffer}, D. and {Hodgkin}, S.~T. and {Holl}, B. and {Jan{\ss}en}, K. and {Jevardat de Fombelle}, G. and {Jordan}, S. and {Krone-Martins}, A. and {Lanzafame}, A.~C. and {L{\"o}ffler}, W. and {Marchal}, O. and {Marrese}, P.~M. and {Moitinho}, A. and {Muinonen}, K. and {Osborne}, P. and {Pancino}, E. and {Pauwels}, T. and {Recio-Blanco}, A. and {Reyl{\'e}}, C. and {Riello}, M. and {Rimoldini}, L. and {Roegiers}, T. and {Rybizki}, J. and {Sarro}, L.~M. and {Siopis}, C. and {Smith}, M. and {Sozzetti}, A. and {Utrilla}, E. and {van Leeuwen}, M. and {Abbas}, U. and {{\'A}brah{\'a}m}, P. and {Abreu Aramburu}, A. and {Aerts}, C. and {Aguado}, J.~J. and {Ajaj}, M. and {Aldea-Montero}, F. and {Altavilla}, G. and {{\'A}lvarez}, M.~A. and {Alves}, J. and {Anders}, F. and {Anderson}, R.~I. and {Anglada Varela}, E. and {Antoja}, T. and {Baines}, D. and {Baker}, S.~G. and {Balaguer-N{\'u}{\~n}ez}, L. and {Balbinot}, E. and {Balog}, Z. and {Barache}, C. and {Barbato}, D. and {Barros}, M. and {Barstow}, M.~A. and {Bartolom{\'e}}, S. and {Bassilana}, J. -L. and {Bauchet}, N. and {Becciani}, U. and {Bellazzini}, M. and {Berihuete}, A. and {Bernet}, M. and {Bertone}, S. and {Bianchi}, L. and {Binnenfeld}, A. and {Blanco-Cuaresma}, S. and {Blazere}, A. and {Boch}, T. and {Bombrun}, A. and {Bossini}, D. and {Bouquillon}, S. and {Bragaglia}, A. and {Bramante}, L. and {Breedt}, E. and {Bressan}, A. and {Brouillet}, N. and {Brugaletta}, E. and {Bucciarelli}, B. and {Burlacu}, A. and {Butkevich}, A.~G. and {Buzzi}, R. and {Caffau}, E. and {Cancelliere}, R. and {Cantat-Gaudin}, T. and {Carballo}, R. and {Carlucci}, T. and {Carnerero}, M.~I. and {Carrasco}, J.~M. and {Casamiquela}, L. and {Castellani}, M. and {Castro-Ginard}, A. and {Chaoul}, L. and {Charlot}, P. and {Chemin}, L. and {Chiaramida}, V. and {Chiavassa}, A. and {Chornay}, N. and {Comoretto}, G. and {Contursi}, G. and {Cooper}, W.~J. and {Cornez}, T. and {Cowell}, S. and {Crifo}, F. and {Cropper}, M. and {Crosta}, M. and {Crowley}, C. and {Dafonte}, C. and {Dapergolas}, A. and {David}, M. and {David}, P. and {de Laverny}, P. and {De Luise}, F. and {De March}, R.},
        title = "{Gaia Data Release 3. Summary of the content and survey properties}",
      journal = {\aap},
         year = 2023,
        month = jun,
       volume = {674},
          eid = {A1},
        pages = {A1},
          doi = {10.1051/0004-6361/202243940},
archivePrefix = {arXiv},
       eprint = {2208.00211},
 primaryClass = {astro-ph.GA},
}

@article{Welsh2013CircumstellarExocomets,
	author = {{Welsh}, Barry Y. and {Montgomery}, Sharon},
        title = "{Circumstellar Gas-Disk Variability Around A-Type Stars: The Detection of Exocomets?}",
      journal = {\pasp},
         year = 2013,
        month = jul,
       volume = {125},
       number = {929},
        pages = {759},
          doi = {10.1086/671757},
       adsurl = {https://ui.adsabs.harvard.edu/abs/2013PASP..125..759W}
}

@article{Welsh2015TheAbsorption,
	author = {{Welsh}, Barry Y. and {Montgomery}, Sharon L.},
        title = "{The Appearance and Disappearance of Exocomet Gas Absorption}",
      journal = {Advances in Astronomy},
         year = 2015,
        month = jan,
       volume = {2015},
          eid = {980323},
        pages = {980323},
          doi = {10.1155/2015/980323},
       adsurl = {https://ui.adsabs.harvard.edu/abs/2015AdAst2015E..26W}
}

@article{Welsh2018FurtherDiscs,
	author = {{Welsh}, Barry Y. and {Montgomery}, Sharon L.},
        title = "{Further detections of exocomet absorbing gas around Southern hemisphere A-type stars with known debris discs}",
      journal = {\mnras},
         year = 2018,
        month = feb,
       volume = {474},
       number = {2},
        pages = {1515-1525},
          doi = {10.1093/mnras/stx2800},
       adsurl = {https://ui.adsabs.harvard.edu/abs/2018MNRAS.474.1515W}
}

@article{Welsh1998Beta85905,
	author = {{Welsh}, B.~Y. and {Craig}, N. and {Crawford}, I.~A. and {Price}, R.~J.},
        title = "{Beta Pic-like circumstellar disk gas surrounding HR 10 and HD 85905}",
      journal = {\aap},
         year = 1998,
        month = oct,
       volume = {338},
        pages = {674-682},
       adsurl = {https://ui.adsabs.harvard.edu/abs/1998A&A...338..674W}
}

@article{Wenger2000TheDatabase,
	author = {{Wenger}, M. and {Ochsenbein}, F. and {Egret}, D. and {Dubois}, P. and {Bonnarel}, F. and {Borde}, S. and {Genova}, F. and {Jasniewicz}, G. and {Lalo{\"e}}, S. and {Lesteven}, S. and {Monier}, R.},
        title = "{The SIMBAD astronomical database. The CDS reference database for astronomical objects}",
      journal = {\aaps},
         year = 2000,
        month = apr,
       volume = {143},
        pages = {9-22},
          doi = {10.1051/aas:2000332},
archivePrefix = {arXiv},
       eprint = {astro-ph/0002110},
 primaryClass = {astro-ph},
       adsurl = {https://ui.adsabs.harvard.edu/abs/2000A&AS..143....9W}
}

@article{Wichittanakom2020TheStars,
	 author = {{Wichittanakom}, C. and {Oudmaijer}, R.~D. and {Fairlamb}, J.~R. and {Mendigut{\'\i}a}, I. and {Vioque}, M. and {Ababakr}, K.~M.},
        title = "{The accretion rates and mechanisms of Herbig Ae/Be stars}",
      journal = {\mnras},
         year = 2020,
        month = mar,
       volume = {493},
       number = {1},
        pages = {234-249},
          doi = {10.1093/mnras/staa169},
archivePrefix = {arXiv},
       eprint = {2001.05971},
 primaryClass = {astro-ph.SR},
       adsurl = {https://ui.adsabs.harvard.edu/abs/2020MNRAS.493..234W}
}

@article{Wilson1994AbundancesMedium,
	author = {{Wilson}, T.~L. and {Rood}, R.},
        title = "{Abundances in the Interstellar Medium}",
      journal = {\araa},
         year = 1994,
        month = jan,
       volume = {32},
        pages = {191-226},
          doi = {10.1146/annurev.aa.32.090194.001203},
       adsurl = {https://ui.adsabs.harvard.edu/abs/1994ARA&A..32..191W}
}

@ARTICLE{Wyatt2002Collisions,
       author = {{Wyatt}, M.~C. and {Dent}, W.~R.~F.},
        title = "{Collisional processes in extrasolar planetesimal discs - dust clumps in Fomalhaut's debris disc}",
      journal = {\mnras},
         year = 2002,
        month = aug,
       volume = {334},
       number = {3},
        pages = {589-607},
          doi = {10.1046/j.1365-8711.2002.05533.x},
archivePrefix = {arXiv},
       eprint = {astro-ph/0204034},
 primaryClass = {astro-ph},
       adsurl = {https://ui.adsabs.harvard.edu/abs/2002MNRAS.334..589W}
}

@article{Wyatt2015FiveDisk,
	 author = {{Wyatt}, M.~C. and {Pani{\'c}}, O. and {Kennedy}, G.~M. and {Matr{\`a}}, L.},
        title = "{Five steps in the evolution from protoplanetary to debris disk}",
      journal = {\apss},
         year = 2015,
        month = jun,
       volume = {357},
       number = {2},
          eid = {103},
        pages = {103},
          doi = {10.1007/s10509-015-2315-6},
archivePrefix = {arXiv},
       eprint = {1412.5598},
 primaryClass = {astro-ph.EP},
       adsurl = {https://ui.adsabs.harvard.edu/abs/2015Ap&SS.357..103W}
}

@article{Zorec2012RotationalStars,
	author = {{Zorec}, J. and {Royer}, F.},
        title = "{Rotational velocities of A-type stars. IV. Evolution of rotational velocities}",
      journal = {\aap},
         year = 2012,
        month = jan,
       volume = {537},
          eid = {A120},
        pages = {A120},
          doi = {10.1051/0004-6361/201117691},
archivePrefix = {arXiv},
       eprint = {1201.2052},
 primaryClass = {astro-ph.SR},
       adsurl = {https://ui.adsabs.harvard.edu/abs/2012A&A...537A.120Z}
}

@ARTICLE{Schneiderman2021,
       author = {{Schneiderman}, Tajana and {Matr{\`a}}, Luca and {Jackson}, Alan P. and {Kennedy}, Grant M. and {Kral}, Quentin and {Marino}, Sebasti{\'a}n and {{\"O}berg}, Karin I. and {Su}, Kate Y.~L. and {Wilner}, David J. and {Wyatt}, Mark C.},
        title = "{Carbon monoxide gas produced by a giant impact in the inner region of a young system}",
      journal = {\nat},
         year = 2021,
        month = oct,
       volume = {598},
       number = {7881},
        pages = {425-428},
          doi = {10.1038/s41586-021-03872-x},
archivePrefix = {arXiv},
       eprint = {2110.15377},
 primaryClass = {astro-ph.EP},
       adsurl = {https://ui.adsabs.harvard.edu/abs/2021Natur.598..425S}
}

@ARTICLE{Beckwith1990,
       author = {{Beckwith}, Steven V.~W. and {Sargent}, Anneila I. and {Chini}, Rolf S. and {Guesten}, Rolf},
        title = "{A Survey for Circumstellar Disks around Young Stellar Objects}",
      journal = {\aj},
         year = 1990,
        month = mar,
       volume = {99},
        pages = {924},
          doi = {10.1086/115385},
       adsurl = {https://ui.adsabs.harvard.edu/abs/1990AJ.....99..924B}
}

@proceedings{PPVII2023,
  title        = {Protostars and Planets VII},
  editor       = {Inutsuka, Shu-ichiro and Aikawa, Yuri and Muto, Takayuki and Tomida, Kengo and Tamura, Motohide},
  booktitle    = {ASP Conference Series, Volume 534},
  year         = {2023},
  publisher    = {Astronomical Society of the Pacific},
  address      = {San Francisco, CA}
}

@ARTICLE{DiFolco2020,
       author = {{Di Folco}, E. and {P{\'e}ricaud}, J. and {Dutrey}, A. and {Augereau}, J. -C. and {Chapillon}, E. and {Guilloteau}, S. and {Pi{\'e}tu}, V. and {Boccaletti}, A.},
        title = "{An ALMA/NOEMA study of gas dissipation and dust evolution in the 5 Myr-old HD 141569A hybrid disc}",
      journal = {\aap},
         year = 2020,
        month = mar,
       volume = {635},
          eid = {A94},
        pages = {A94},
          doi = {10.1051/0004-6361/201732243},
       adsurl = {https://ui.adsabs.harvard.edu/abs/2020A&A...635A..94D}
}

@article{Matra2018AnRelation,
    title = {{An Empirical Planetesimal Belt Radius–Stellar Luminosity Relation}},
    year = {2018},
    journal = {The Astrophysical Journal},
    author = {Matr{\`{a}}, L. and Marino, S. and Kennedy, G. M. and Wyatt, M. C. and {\"{O}}berg, K. I. and Wilner, D. J.},
    number = {1},
    month = {5},
    pages = {72},
    volume = {859},
    doi = {10.3847/1538-4357/aabcc4},
    issn = {1538-4357}
}

@article{Rhee2007CharacterizationCatalogs,
    title = {{Characterization of Dusty Debris Disks: The IRAS and Hipparcos Catalogs}},
    year = {2007},
    journal = {The Astrophysical Journal},
    author = {Rhee, Joseph H. and Song, Inseok and Zuckerman, B. and McElwain, Michael},
    number = {2},
    month = {5},
    pages = {1556--1571},
    volume = {660},
    doi = {10.1086/509912},
    issn = {0004-637X}
}

@ARTICLE{Lebreton2013,
       author = {{Lebreton}, J. and {van Lieshout}, R. and {Augereau}, J. -C. and {Absil}, O. and {Mennesson}, B. and {Kama}, M. and {Dominik}, C. and {Bonsor}, A. and {Vandeportal}, J. and {Beust}, H. and {Defr{\`e}re}, D. and {Ertel}, S. and {Faramaz}, V. and {Hinz}, P. and {Kral}, Q. and {Lagrange}, A. -M. and {Liu}, W. and {Th{\'e}bault}, P.},
        title = "{An interferometric study of the Fomalhaut inner debris disk. III. Detailed models of the exozodiacal disk and its origin}",
      journal = {\aap},
         year = 2013,
        month = jul,
       volume = {555},
          eid = {A146},
        pages = {A146},
archivePrefix = {arXiv},
       eprint = {1306.0956},
 primaryClass = {astro-ph.EP},
}

@ARTICLE{Pearce2020,
       author = {{Pearce}, Tim D. and {Krivov}, Alexander V. and {Booth}, Mark},
        title = "{Gas trapping of hot dust around main-sequence stars}",
      journal = {\mnras},
         year = 2020,
        month = oct,
       volume = {498},
       number = {2},
        pages = {2798-2813},
archivePrefix = {arXiv},
       eprint = {2008.07505},
 primaryClass = {astro-ph.EP},
}

@ARTICLE{Sezestre2019,
       author = {{Sezestre}, {\'E}. and {Augereau}, J. -C. and {Th{\'e}bault}, P.},
        title = "{Hot exozodiacal dust: an exocometary origin?}",
      journal = {\aap},
         year = 2019,
        month = jun,
       volume = {626},
          eid = {A2},
        pages = {A2}
}

@ARTICLE{Squicciarini2025,
       author = {{Squicciarini}, V. and {Mazoyer}, J. and {Lagrange}, A. -M. and {Chomez}, A. and {Delorme}, P. and {Flasseur}, O. and {Kiefer}, F. and {Bergeon}, S. and {Albert}, D. and {Meunier}, N.},
        title = "{The COBREX archival survey: Improved constraints on the occurrence rate of wide-orbit substellar companions: I. A uniform re-analysis of 400 stars from the GPIES survey}",
      journal = {\aap},
         year = 2025,
        month = jan,
       volume = {693},
          eid = {A54},
        pages = {A54},
          doi = {10.1051/0004-6361/202452310},
archivePrefix = {arXiv},
       eprint = {2411.06157},
 primaryClass = {astro-ph.EP},
       adsurl = {https://ui.adsabs.harvard.edu/abs/2025A&A...693A..54S}
}

@ARTICLE{Fairlamb2015,
       author = {{Fairlamb}, J.~R. and {Oudmaijer}, R.~D. and {Mendigut{\'\i}a}, I. and {Ilee}, J.~D. and {van den Ancker}, M.~E.},
        title = "{A spectroscopic survey of Herbig Ae/Be stars with X-shooter - I. Stellar parameters and accretion rates}",
      journal = {\mnras},
         year = 2015,
        month = oct,
       volume = {453},
       number = {1},
        pages = {976-1001},
}

@ARTICLE{Borthakur2025,
       author = {{Borthakur}, Sandipan P.~D. and {Kama}, Mihkel and {Fossati}, Luca and {Kral}, Quentin and {Folsom}, Colin P. and {Teske}, Johanna and {Aret}, Anna},
        title = "{Abundance analysis of stars hosting gas-rich debris discs}",
      journal = {\aap},
         year = 2025,
        month = may,
       volume = {697},
          eid = {A59},
        pages = {A59}
}

%\begin{thebibliography}{99}
%\end{thebibliography}

%%%%%%%%%%%%%%%%%%%%%%%%%%%%%%%%%%%%%%%%%%%%%%%%%%

%%%%%%%%%%%%%%%%% APPENDICES %%%%%%%%%%%%%%%%%%%%%

\appendix

\section{Debris discs with gas found through emission}

\clearpage

\begin{sidewaystable}
\centering
\captionsetup{width=0.9\textwidth}
\begin{tabular}{lllllllll}
\hline
Name & Age [Myr] & Stellar Mass [$M_{\odot}$] & Spectral Type & Distance {[}pc{]} & Dust mass [$M_{\oplus}$] & CO gas mass [$M_{\oplus}$] & Reference \\ \hline
V*\,NO\,Lup & 1-3 & 0.7 & K7 & 132.90 & 4.00\,$\times$\,10$^{-2}$ & 5.03\,$\times$\,10$^{-5}$ & \citet{Lovell2021ALMADispersal}$^{e}$ \\ \hline
HD\,44892 & 2.1 & 2.87 & A9/F0IV & 191.00 & 1.91\,$\times$\,10$^{-2}$ & 1.62\,$\times$\,10$^{-4}$ & This study \\ \hline
HD\,141569 & 5 & 2.04$^{[1]}$ & A2VekB9mB9(\_lB) & 111.61  & 2.00\,$\times$\,10$^{-1}$ & 3.01\,$\times$\,10$^{-1}$ & \makecell[l]{\citet{Miley2018Unlocking141569}$^{e}$ \\ \citet{Malamut2014THEPc}$^{a}$} \\ \hline
HD\,36546 & 3-10 & 1.87$^{[2]}$ & B8 & 100.18 & 7.40\,$\times$\,10$^{-2}$ & 6.69\,$\times$\,10$^{-3}$ & \makecell[l]{ \citet{Rebollido2022The36546}$^{e}$\\ \citet{Rebollido2020Exocomets:Survey}$^{a}$} \\ \hline
V*\,CE\,Ant & 10 & 0.31$^{[1]}$ & M2Ve & 34.10 & 2.13\,$\times$\,10$^{-2}$ & 1.68\,$\times$\,10$^{-6}$ & \citet{Matra2019On7}$^{e}$ \\ \hline
HD\,138813 & 10$^{[3]}$ & 2.15$^{[3]}$ & A0V & 136.60 & 1.30\,$\times$\,10$^{-1}$ & 1.23\,$\times$\,10$^{-3}$ & \citet{Lieman-Sifry2016DebrisALMA}$^{e}$ \\ \hline
HD\,146897 & 10$^{[3]}$ & 1.32$^{[3]}$ & F2/3V & 132.19 & 2.14\,$\times$\,10$^{-1}$ & 2.90\,$\times$\,10$^{-6}$ & \citet{Lieman-Sifry2016DebrisALMA}$^{e}$ \\ \hline
HD\,172555 & 24$^{[3]}$ & 1.71$^{[3]}$ & A7V & 28.79 & 1.79\,$\times$\,10$^{-4}$  & 4.56\,$\times$\,10$^{-6}$ & \makecell[l]{\citet{Schneiderman2021}$^{e}$ \\  \citet{Kiefer2014ExocometsHD172555}$^{a}$} \\ \hline
HD\,110058 & 15$^{[3]}$ & 1.53$^{[3]}$ & A0V & 130.08 & 3.23\,$\times$\,10$^{-2}$ & 5.40\,$\times$\,10$^{-3}$ & \makecell[l]{ \citet{Hales2022ALMADisk}$^{e}$ \\ \citet{Iglesias2018DebrisOrigin}$^{a}$} \\ \hline
HD\,121191 & 16$^{[3]}$ & 1.63$^{[3]}$ & A5IV/V & 132.29 & 9.02\,$\times$\,10$^{-3}$ & 5.83\,$\times$\,10$^{-3}$ & \citet{Moor2017MolecularStars}$^{e}$ \\ \hline
HD\,121617 & 16$^{[3]}$ & 1.90$^{[3]}$ & A1V & 117.89 & 1.30\,$\times$\,10$^{-1}$ & 3.86\,$\times$\,10$^{-2}$ & \citet{Moor2017MolecularStars}$^{e}$ \\ \hline
HD\,131488 & 16$^{[4]}$ & 1.8$^{[6]}$ & A1V & 152.24 & 3.21\,$\times$\,10$^{-1}$ & 4.75\,$\times$\,10$^{-2}$ & \citet{Moor2017MolecularStars}$^{e}$ \\ \hline
HD\,131835 & 16$^{[3]}$ & 1.81$^{[3]}$ & A2IV & 129.74 & 2.81\,$\times$\,10$^{-1}$ & 2.14\,$\times$\,10$^{-2}$ & \citet{Lieman-Sifry2016DebrisALMA}$^{e}$ \\ \hline
HD\,156623 & 16$^{[3]}$ & 1.91$^{[3]}$ & A0V & 108.33 & 3.62\,$\times$\,10$^{-2}$ & 6.48\,$\times$\,10$^{-4}$ & \makecell[l]{\citet{Lieman-Sifry2016DebrisALMA}$^{e}$ \\ \citet{Rebollido2018TheDiscs}$^{a}$} \\ \hline
%HD\,95086 & 17 & 1.61$^{[1]}$ & A8III & 86.46 & 3.96\,$\times$\,10$^{-1}$ & 3.32\,$\times$\,10$^{-6}$ & \citet{Booth2019DeepDisc}$^{e}$ \\ \hline
HD\,181327 & 24 & 1.36 & F6V & 47.78 & 2.41\,$\times$\,10$^{-1}$ & 2.70\,$\times$\,10$^{-6}$ & \citet{Marino2016ExocometaryRing}$^{e}$ \\ \hline
*\,bet\,Pic & 24$^{[7]}$ & 1.73$^{[1]}$ & A6V & 19.63 & 4.13\,$\times$\,10$^{-2}$ & 2.41\,$\times$\,10$^{-5}$ & \makecell[l]{\citet{Dent2014MolecularDisk}$^{e}$ \\ \citet{Kiefer2014TwoSystem}$^{a}$} \\ \hline
%HD\,197481 & 24$^{[3]}$ & 0.59$^{[3]}$ & M1VeBa1 & 9.71 & - & - & \citet{France2007ADisk}$^{e}$ \\ \hline
*\,eta\,Tel & 24$^{[3]}$ & 2.21$^{[3]}$ & A0V & 48.54 & 1.33\,$\times$\,10$^{-2}$ & - & \makecell[l]{\citet{Riviere-Marichalar2014GasObservatory}$^{e}$ \\ \citet{Rebollido2018TheDiscs}$^{a}$} \\ \hline
HD\,32297 & <30 & 1.69$^{[1]}$ & A0V & 129.73 & 6.04\,$\times$\,10$^{-1}$ & 1.56\,$\times$\,10$^{-1}$ & \makecell[l]{\citet{Moor2019New32297}$^{e}$ \\ \citet{Redfield2007Gas32297}$^{a}$} \\ \hline
*\,49\,Cet & 45$^{[3]}$ & 1.98$^{[3]}$ & A1V & 57.23 & 1.44\,$\times$\,10$^{-1}$ & 2.42\,$\times$\,10$^{-2}$ & \makecell[l]{ \citet{Moor2019New32297}$^{e}$ \\ \citet{Roberge2014VOLATILE-RICHDISK}$^{a}$} \\ \hline
HD\,21997 & 42$^{[3]}$ & 1.83$^{[3]}$ & A3IV/V & 69.69 & 2.64\,$\times$\,10$^{-1}$ & 6.17\,$\times$\,10$^{-2}$ & \citet{Kospal2013ALMA21997}$^{e}$ \\ \hline
V*\,V1229\,Tau & 170$^{[5]}$ & - & A0VpSi+Am & 138.31 & - & 6.33\,$\times$\,10$^{-4}$ & \citet{Pericaud2017TheDisks}$^{e}$ \\ \hline
HD\,216956 & 440 & 1.9$^{[6]}$ & A4V & 7.70 & 1.49\,$\times$\,10$^{-2}$ & 1.89\,$\times$\,10$^{-7}$ & \citet{Matra2017DetectionComets}$^{e}$\\ \hline
*\,eta\,Crv & 1000-2000 & 1.39$^{[3]}$ & F2V & 18.24 & 1.07\,$\times$\,10$^{-2}$ & 5.92\,$\times$\,10$^{-7}$ & \citet{Marino2017ALMAPlanets}$^{e}$ \\ \hline
\end{tabular}
\caption{$^{e}$ denotes references for gas emission detections, while $^{a}$ refers to gas absorption discoveries in discs that also exhibit gas emission. Dust and CO gas masses were uniformly calculated across the sample. For CO gas mass, we assumed a uniform gas temperature T\,=\,50\,K. Where available, the CO gas mass estimates are based on observations of rarer isotopologues. For dust mass, we used dust temperatures derived for corresponding targets, with dust opacity assumed as $\kappa=0.1\times2.3\times(\frac{\nu}{230\times10^9})^{0.7}$ [cm$^{2}$\,g$^{-1}$] \citep{Beckwith1990}, where $\nu$ is the frequency of observations [Hz]. Stellar ages and masses are adopted from the listed references unless indicated otherwise. Spectral types and distances were obtained from SIMBAD \citep{Wenger2000TheDatabase}. We do not report CO gas mass for *\,eta\,Tel, which has [CII] emission. For V*\,V1229 Tau no continuum emission was detected, hence we do not report the dust mass in this disc. We did not find a reliable stellar mass estimate for V\,*\,V1229 Tau. Gas absorption detections and their references were adopted from \citet{Iglesias2020SearchingDisks}, Table\,1.1. Additional references for stellar age and mass: [1]\,\citet{Esposito2020DebrisCampaign}, [2]\,\citet{AllendePrieto1999FundamentalTemperatures}, [3]\,\citet{Pearce2022PlanetDiscs}, [4]\,\citet{Moor2017MolecularStars}, [5]\,\citet{Southworth2022ScutiHD23642}, [6]\,\citet{Matra2018AnRelation},
[7]\,\citet{Squicciarini2025}.}
\label{tab:allCOGasDetections}
\end{sidewaystable}

%%%%%%%%%%%%%%%%%%%%%%%%%%%%%%%%%%%%%%%%%%%%%%%%%%

% Don't change these lines
\bsp	% typesetting comment
\label{lastpage}
\end{document}